\documentclass{aa}  

\usepackage{graphicx}
\usepackage{txfonts}
\usepackage{placeins}           
                                
\usepackage{multirow}
\usepackage{booktabs}
\usepackage{siunitx}

\usepackage{hyperref}

\begin{document}

   \title{The atomic multiphase interstellar medium of galaxies in the COLIBRE simulations}
   \titlerunning{The multiphase ISM in COLIBRE galaxies}
   \authorrunning{M. Huppenkothen et al.}

%
%
%

   \author{Moa Huppenkothen\inst{\ref{uniwien}}\corrauth{a11917965@unet.univie.ac.at}        
        \and Sylvia Ploeckinger\inst{\ref{uniwien}}\corrauth{sylvia.ploeckinger@univie.ac.at} 
        \and Joop Schaye\inst{\ref{leiden}}\email{schaye@strw.leidenuniv.nl}
        \and Alejandro Benítez-Llambay\inst{\ref{ABLinst}}\email{alejandro.benitezllambay@unimib.it}
        \and Evgenii Chaikin\inst{\ref{ECinst},\ref{leiden}}\email{evgenii.chaikin@durham.ac.uk}
        \and Filip Huško\inst{\ref{leiden}}\email{husko@strw.leidenuniv.nl} 
        \and Alexander J. Richings\inst{\ref{ARinst1},\ref{ARinst2}}\email{A.J.Richings@hull.ac.uk}
        \and James W. Trayford\inst{\ref{JTinst}}\email{j.trayford@herts.ac.uk}
        \and Alvaro Hacar\inst{\ref{uniwien}}\email{alvaro.hacar@univie.ac.at}
        \and Juan D. Soler\inst{\ref{uniwien}}\email{juandiegosolerp@gmail.com}
        }

   \institute{
    Department of Astrophysics, University of Vienna, T\"urkenschanzstrasse 17, 1180 Vienna, Austria \label{uniwien} 
    \and Leiden Observatory, Leiden University, PO Box 9513, 2300 RA Leiden, the Netherlands \label{leiden} 
    \and Dipartimento di Fisica "Giuseppe Occhialini", Università degli Studi di Milano-Bicocca \label{ABLinst} 
    \and Institute for Computational Cosmology, Department of Physics, University of Durham, South Road, Durham, DH1 3LE, UK \label{ECinst}     
    \and Centre for Data Science, Artificial Intelligence and Modelling, University of Hull, Cottingham Road, Hull, HU6 7RX, UK \label{ARinst1} 
    \and E. A. Milne Centre for Astrophysics, University of Hull, Cottingham Road, Hull, HU6 7RX, UK \label{ARinst2} 
    \and Centre for Astrophysics Research, Department of Physics, Astronomy and Mathematics, University of Hertfordshire, College Lane, Hatfield, AL10 9AB, UK \label{JTinst}
   }

   \date{Received 10 September 2026 / Accepted date}

 
  \abstract
   {COLIBRE is the first cosmological simulation project that directly models the multiphase interstellar medium (ISM) without a pressure floor and including a live-dust model. Its radiative cooling model tracks hydrogen and helium in non-equilibrium and is connected to the live-dust model. 
   This facilitates an in-depth and self-consistent analysis of the neutral gas within galaxies: the warm ($T\gtrsim1000\,\mathrm{K}$) and cold ($T\lesssim1000\,\mathrm{K}$) neutral media.}
   {We analyse the neutral ISM of galaxies at redshift $z=0$ in the COLIBRE hydrodynamical simulations and investigate the conditions under which these two gas phases coexist in a narrow range of thermal pressures.} 
   {COLIBRE galaxies are selected based on the metallicity of their ISM. The median and mass-weighted distributions of thermal pressures of the multiphase ISM are analysed and compared to thermal equilibrium models and observations.}
   {The ISM in galaxies with gas metallicities similar to solar values exhibits a clear multiphase structure with a warm and cold phase coexisting in a certain range of thermal pressures. The pressures at which the ISM is multiphase depend on the gas metallicity. For COLIBRE galaxies with lower metallicities ($Z_{\mathrm{ISM}}\lesssim0.1\,\mathrm{Z}_{\odot}$), this multiphase structure largely disappears, partly due to resolution. The thermal pressures weighted by the \ion{H}{I} mass of the neutral phases in COLIBRE galaxies are lower than, but still comparable to, some theoretical works and observational estimates. The thermal pressures show a strong dependence on the weighting scheme. 
   If weighted by the star formation rate or \ion{C}{I} mass, the thermal pressures of the cold phase match those derived from observations.}
   {The resulting pressures depend on a combination of the assumed radiation field strength, dust abundance, limited resolution and the weighting scheme. The strong dependence of the thermal pressures on the weighting scheme indicates observational tracers used to estimate the thermal pressure are biased towards high-pressure regions.}


   \maketitle
    \nolinenumbers


    \begin{table*} 
        \caption{\textsc{colibre} simulations analysed in this work. The identifier (first column) denotes both the side length of the simulated volume ($L$; second column), and the resolution level, which is specified by the initial average baryonic particle mass ($m_{\mathrm{g}}$; third column), the gravitational force softening length at $z=0$ ($\epsilon$; fourth column), the normalization of the interstellar radiation field ($R_{\rm ISRF}$; fourth column) used in the cooling model (see \protect\citealp{P25}).
        The final column lists the sections that discuss the results of these simulations. }
        \centering
        \begin{tabular}{llllll}
            \toprule
            \multirow{2}{*}{Identifier} & $L$ & $m_{\mathrm{g}}$ & $\epsilon$ & \multirow{2}{*}{$R_{\rm ISRF}$} & \multirow{2}{*}{Section}\\
             & [cMpc] & [$\mathrm{M}_{\odot}$] & [kpc] &  &  \\
            \midrule
            L025m5 &  25 & $2.30\times10^{5}$ & 0.35 & 0.1 & \ref{sec:MW_analogue}, \ref{sec:resolution} \\
            L200m6 & 200 & $1.84\times10^{6}$ & 0.7 & 0.1 & \ref{sec:metallicity_dependence}, \ref{sec:resolution} \\
            L025m6 &  25 & $1.84\times10^{6}$ & 0.7 & 0.1 & \ref{sec:resolution} \\
            L025m6 ISRF &  25 & $1.84\times10^{6}$ & 0.7 & 1 & \ref{sec:disc:weakISRF}, \ref{sec:disc:theory} \\  
            L025m7 &  25 & $1.47\times10^{7}$ & 1.4 & 0.1 & \ref{sec:resolution} \\
            \bottomrule
        \end{tabular}
        \label{tab:colibre_volumes}
    \end{table*}

\section{Introduction}

The interstellar medium (ISM) consists of molecular, atomic, and ionized gas as well as dust grains, which interact with the interstellar radiation field (ISRF) and cosmic rays (CRs). As a result, the ISM exhibits a complex structure regulated by gas heating and cooling, turbulence, molecule formation, and stellar feedback (e.g.~\citealp{Draine2011, Klessen2016}). Despite this complexity, robust physical trends emerge. Neutral hydrogen measurements indicate that cold, dense structures with temperatures of $T \sim 100\,\mathrm{K}$ are frequently embedded within more extended, warm, diffuse atomic gas at $T \approx 6\,000\, - \,10\,000\,\mathrm{K}$ \citep{McKeeOstriker1977, KalberlaHaud2018}. The cold neutral medium (CNM) is observationally identified through narrow 21-cm absorption features, whereas the surrounding warm neutral medium (WNM) appears as broad 21-cm emission-line envelopes (e.g.~\citealp{Dickey1978, Dickey1990, Heiles2003, KalberlaKerp2009}). These phases can coexist in approximate thermal balance and near pressure equilibrium, giving rise to the canonical two-phase model of the atomic ISM from \citet{Field1969}, well separated by a thermal instability \citep{Field1965}. 

The physical properties (i.e. volume density, temperature, pressure) at which warm and cold gas can coexist have been evaluated by comparing radiative heating and cooling rates for various assumptions on radiation fields, shielding column densities, cosmic ray rates, and gas metallicities \citep{Wolfire1995, Wolfire2003, Bialy2019, Kim2023}. For assumptions on the ISRF, CR rate, and metallicity, that are representative of the solar neighbourhood, the expected thermal pressures match observational estimates of the local CNM (e.g. from \citealp{JT11}). 

The multiphase structure of the ISM has been widely studied with hydrodynamic simulations of turbulent colliding flows (e.g.~\citealp{Audit2005}, \mbox{\citealp{Vazquez2006}}, \citealp{Saury2014}), supernova-driven stratified boxes (e.g.~\citealp{deAvillez2004,  Walch2015, Kim2017Tigress, Iffrig2017, Vijayan2020}), as well as with simulations of isolated disks (e.g.~\citealp{Richings2016, Benincasa2016, HuChia2017, Deng2024}) or galactic-scale cosmological zoom-in simulations (e.g.~\citealp{Gurvich2020}). 

Until recently, cosmological simulations of large ($\gtrsim 100^3\ \mathrm{Mpc}^3$) volumes have effectively modelled the ISM as a single phase, following a polytropic equation of state or applying a temperature, pressure, or entropy floor for gas with densities of $n_{\mathrm{H}}\gtrsim0.1\,\mathrm{cm}^{-3}$ (e.g. \textsc{owls}: \citealp{Schaye2010owls}; \textsc{horizon-agn}: \citealp{Dubois2014}; \textsc{eagle}: \citealp{Schaye2015}; IllustrisTNG: \citealp{Pillepich2018}; \textsc{simba}: \citealp{Dave2019}; \textsc{astrid}: \citealp{Bird2022, Zhou2026}; \textsc{crocodile}: \citealp{Oku2024}). 
This approach has often been motivated by the requirement to resolve the Jeans mass \citep{Jeans1902} of self-gravitating structures to avoid artificial fragmentation. However, \citet{Ploeckinger2024} demonstrate that in softened gravity (as is typical for Lagrangian simulations), artificial fragmentation is not a concern.
Indeed, in recent years, a few simulation projects with a multiphase ISM in small ($< 25^{3}\,\rm Mpc^{3}$) cosmological volumes have been realized, such as \textsc{Newhorizon} \citep{Dubois2021} and \textsc{Firebox} \citep{Feldmann2023}.

The \textsc{colibre} project \citep{Schaye2025colibre, Chaikin2025calibration} is the first cosmological simulation project of large volumes (e.g. a cube with a side length of 200 comoving Mpc with a mass per resolution element of $\approx 10^6\,\mathrm{M}_{\odot}$) evolved until $z=0$ that directly models the multiphase ISM and the evolution of dust grains. 
We analyse the multiphase structure of the atomic ISM in \textsc{colibre} galaxies at $z=0$, focusing on the thermal pressure range where the warm and cold phases can coexist. 

The paper is organized as follows. In Sect.~\ref{sec:method}, we outline the datasets and methods used in this work. We review the key aspects of the \textsc{colibre} simulations in Sect.~\ref{sec:colibre}, focusing on the ingredients of the neutral gas model (Sect.~\ref{sec:hybrid-chimes}) and the relevant galaxy properties from the \textsc{colibre} catalogues (Sect.~\ref{sec:galaxy_properties}). 
We analyse the multiphase ISM of \textsc{colibre} galaxies in Sect.~\ref{sec:results}, in which we first highlight four individual galaxies with approximately solar metallicity (Sect.~\ref{sec:MW_analogue}), and in Sect.~\ref{sec:metallicity_dependence}, we assess the metallicity dependence of the multiphase ISM for the more general \textsc{colibre} galaxy population. 
In Sect.~\ref{sec:discussion} we discuss the results in the context of the assumed strength of the ISRF in \textsc{colibre} (Sect.~\ref{sec:disc:weakISRF}), the emerging dust-to-gas ratio from the live-dust model (Sect.~\ref{sec:disc:dust}), and the resolution (Sect.~\ref{sec:resolution}). In Sect.~\ref{sec:discussion_lit}, we compare the results to theoretical (Sect.~\ref{sec:disc:theory}) and observational (Sect.~\ref{sec:disc:obs}) studies, and we summarize our findings in Sect.~\ref{sec:summary}. 

Throughout this work, we refer to steady-state chemistry and ionization equilibrium as ``chemical equilibrium'' or ``equilibrium chemistry'' for brevity. We use the term (chemical) elements to refer to H, He, C, N, O, Ne, Si, Mg, S,
Ca, and Fe, while (chemical) species reference the individual atoms, ions, and molecules, e.g. H$_2$, \ion{H}{I}, \ion{H}{II}, \ion{C}{II}, as well as free electrons. Abundances are defined relative to the total hydrogen number density, $n_{\mathrm{H}}$; for example, the element abundance of C is $n_{\mathrm{C}}/n_{\mathrm{H}}$ and the species abundance of \ion{C}{II} is $n_{\mathrm{CII}}/n_{\mathrm{H}}$. Furthermore, the species fraction of e.g. \ion{C}{II} is the fraction of C atoms that is singly ionized, $n_{\mathrm{CII}}/n_{\mathrm{C}}$. We use solar abundance ratios and the solar metallicity $Z_{\odot} = 0.0134$ from \cite{Asplund2009}. 
The \ion{C}{I} and \ion{C}{II} masses of gas particles in the simulations are calculated from the carbon mass fractions in the gas-phase using \ion{C}{I} and \ion{C}{II} species fractions from the pre-tabulated abundances, which assume chemical equilibrium, at the corresponding metallicity and temperatures and densities of the particles. This is not necessary for the \ion{H}{I} species fractions, since hydrogen and helium species are followed in non-equilibrium chemistry at runtime (see Sect.~\ref{sec:hybrid-chimes}). Finally, we use $\log$ as a shorthand for $\log_{10}$.

\section{Method}\label{sec:method}

\subsection{The COLIBRE simulations}\label{sec:colibre} 

    For this work, we use simulations from the \textsc{colibre} project \citep{Schaye2025colibre, Chaikin2025calibration}.
    We primarily analyse the largest volumes at $z=0$ (25 and 200~cMpc) at the two highest resolutions, m5 and m6, with both baryonic and dark matter particle masses of the order of $10^{5}\,\mathrm{M}_{\odot}$ and $10^{6}\,\mathrm{M}_{\odot}$, respectively (Table~\ref{tab:colibre_volumes}). 

    The \textsc{colibre} simulations are run with the \textsc{swift} \citep{Schaller2024swift} code and the hydrodynamic equations are solved with the density-energy smoothed particle hydrodynamics (SPH) scheme \textsc{sphenix} \citep{Borrow2022}. The initial conditions (ICs) for these cosmological simulations are generated at a redshift of $z = 63$ with \textsc{monofonIC}  \citep{Hahn2020, Michaux2021} for the $\Lambda$CDM cosmology from the Dark Energy Survey \citep{Abbott2022}. When creating the ICs, dark matter is super-sampled compared to the baryons by using four times more dark matter than baryonic particles. The particles representing gas, stars, and dark matter then have comparable masses, which reduces spurious heating of stellar disks by more massive dark matter particles \citep{Ludlow2019, Ludlow2023}. 

    \citet{Chaikin2025calibration} detail the calibration of the subgrid feedback model in \textsc{colibre}, where four parameters (affecting AGN and stellar feedback) are sampled and then fit to the redshift $z = 0$ galaxy stellar mass function \citep{Driver2022} and the $z \approx 0$ size–mass relation \citep{Hardwick2022}, using emulators trained on $\approx 200$ L050m7 simulations. When moving to higher resolution simulations, a select few parameters (see Table~1 of \citealp{Schaye2025colibre}) are slightly adjusted in small-volume test simulations (e.g. L025m6). 
    The simulations match well with observational measurements not used in the calibration. 
    For example, \textsc{colibre} reproduces the $z=0$ trends and evolution through cosmic time of the galaxy stellar mass function and star formation rates \citep{Chaikin2025}, galaxy size–mass relation \citep{Ludlow2026}, gas-phase mass-metallicity relation \citep{Sharda2026}, as well as the spatially resolved Kennicut-Schmidt (KS) relation and the dependence of its scatter on metallicity \citep{Lagos2026KSrelation}. \textsc{colibre} also reproduces the atomic and molecular hydrogen masses, dust masses, and gas and stellar metallicities of $z=0$ galaxies \citep{Schaye2025colibre}. 

    The physical and chemical processes within the ISM in \textsc{colibre} are critical for this work and are therefore laid out in more detail in Sect.~\ref{sec:hybrid-chimes}, while we briefly summarize other key astrophysical process included in \textsc{colibre} here. We refer to \citet{Schaye2025colibre} for more information on the \textsc{colibre} galaxy formation model. Star formation is implemented stochastically, using a Schmidt law and limited to locally gravitationally unstable gas following \citet{Nobels2024}, accounting for both the thermal and turbulent velocity dispersion. Most importantly, while colder gas in this model is more likely to form stars because of the reduced thermal velocity dispersion and its typically higher gas density, no temperature ceiling (or density floor) is imposed when selecting gas particles to be eligible for transforming into a star particle. While star formation in \textsc{colibre} typically occurs in cold ($T<1000\,\mathrm{K}$) gas, it extends to warm ($T\gtrsim1000\,\mathrm{K}$) gas for low gas metallicities and low numerical resolution.

    Following stellar evolution models, star particles enrich the surrounding gas with metals, as described in \cite{Correa2026}, and massive stars inject energy and momentum from ionizing radiation, radiation pressure, and stellar winds, which is modelled as in \cite{BenitezLlambay2026}. Supernova feedback is implemented using both a stochastic thermal and a lower-energy kinetic component, as described in \citet{Chaikin2023}. We use the fiducial \textsc{colibre} runs with purely thermally driven AGN feedback, based on \citet{Booth2009}.

    \begin{table*}
        \centering 
        \caption{Properties of the four selected galaxies (A, B, C, D) from the L025m5 simulation and of the Milky Way Galaxy (MW) for reference. The metallicity $Z_\mathrm{ISM}$ in solar units (second column), dust-to-metal and dust-to-gas mass ratios $\mathcal{DTZ}\ \&\ \mathcal{DTG}$ (third and fourth columns), halo mass $M_\mathrm{halo}$ (fifth column), stellar mass $M_\star$ (sixth), total gas mass $M_\mathrm{Gas}$ (seventh column), total atomic mass $M_\mathrm{HI}$ (eighth column), total molecular mass $M_\mathrm{H_2}$ (ninth column) and SFR (tenth column) for each galaxy (identifier in the first column) evaluated for all particles within a 3D aperture of radius 50 pkpc. The final column lists the unique identifier of the galaxy within the simulation.} 
        \label{tab:MW_ex_properties}  
        \begin{tabular}{c *{9}{S[table-column-width=1.2cm]} c}
            \toprule
            \multirow{2}{*}{\textbf{ID}} & $Z_\mathrm{ISM}$ & $\mathcal{DTZ}$ & $\mathcal{DTG}$ & $M_\mathrm{halo}$ & $M_\star$ & $M_\mathrm{Gas}$ & $M_\mathrm{HI}$ & $M_\mathrm{H_2}$ & SFR & ID\\
             & $[Z_\odot]$ &  & {[10$^{-3}$]} & {[10$^{11}$ M$_\odot$]} & {[10$^{10}$ M$_\odot$]} & {[10$^{10}$ M$_\odot$]} & {[10$^{9}$ M$_\odot$]} & {[10$^{9}$ M$_\odot$]} & {[M$_\odot$ yr$^{-1}$]} &\\
            \midrule
            \textbf{A} & 0.96 & 0.30 & 4.01 & 3.21 & 0.93 & 1.20 & 8.28 & 2.33 & 2.38 & 19191 \\
            \textbf{B} & 1.21 & 0.31 & 4.94 & 9.03 & 1.68 & 0.97 & 8.51 & 1.64 & 1.41 & 24596 \\
            \textbf{C} & 0.92 & 0.27 & 3.36 & 6.34 & 1.51 & 1.09 & 13.34 & 1.20 & 1.13 & 50666 \\
            \textbf{D} & 1.12 & 0.30 & 4.73 & 8.98 & 1.72 & 1.44 & 8.84 & 3.36 & 4.30 & 31393\\
            \midrule
            \textbf{MW} & 1.00$^{\rm a}$ & 0.30$^{\rm b}$ & 3.83$^{\rm b}$ & 9.7$^{\rm c}$ & 5.00$^{\rm c}$ & 1.25$^{\rm d}$ &8.0$^{\rm d}$ &2.5$^{\rm d}$ &1.65$^{\rm e}$ &\\
            \bottomrule
        \end{tabular}
        \tablebib{
            $^{\mathrm{a}}$\citet{Asplund2009}; $^{\mathrm{b}}$\citet{RomanDuval2022}: For a hydrogen column density of $\log N_\mathrm{H}\, [\mathrm{cm}^{-2}] = 20$; $^{\mathrm{c}}$\citet{Cautun2020}; $^{\mathrm{d}}$\citet{KalberlaKerp2009}; $^{\mathrm{e}}$\citet{Licquia2015}
        }
    \end{table*}

\subsubsection{Modelling the multiphase ISM}\label{sec:hybrid-chimes} 
    The ISM model in \textsc{colibre} is described in detail by \citeauthor{P25} (\citeyear{P25}; gas chemistry and radiative cooling) and \citeauthor{Trayford2025} (\citeyear{Trayford2025}; dust model). Because the results of this work need to be interpreted within the context of the various model assumptions, we present the relevant aspects in this section. 

    The multiphase nature of the neutral ISM emerges from a complex network of intertwined physical and chemical processes. Within a galaxy, radiation from both external (distant galaxies and quasars) and internal (local interstellar radiation field, ISRF) sources may ionize atoms, dissociate molecules, and heat the gas. Neutral gas in the ISM is typically partly shielded from ionizing (and dissociating) radiation, which reduces the radiative heating rates. Gas may cool radiatively via collisional excitation and ionization, as well as recombination, for which the cooling rates for each element depend on the respective element species fractions. The gas and ISRF interact with dust grains and vice versa: dust grains efficiently absorb far-UV radiation and therefore provide an important shielding contribution. The absorbed radiation in turn results in the photoelectric (PE) heating of the gas. Finally, dust grains are a critical catalyst for several chemical reactions, most notably the formation of molecular hydrogen. 

    For simulations as computationally expensive as the \textsc{colibre} flagship simulations, including individual sources of radiation, full radiative transfer, and a complete chemical network that solves for the non-equilibrium species abundances on-the-fly is currently unachievable. Nevertheless, \textsc{colibre} includes all critical gas and dust processes for cold gas down to $\approx 10\,\mathrm{K}$, albeit in approximate form.  

    The radiation sources in \textsc{colibre} are a combination of a redshift-dependent metagalactic UV/X-ray background and a more local ISRF. The spectral shape of the ISRF is assumed to be constant and is consistent with measurements at the position of the Sun \citep{Mathis1983,Bregman1986}, while its intensity depends on local gas properties. Following \citet{P25}, this means that higher-density gas, which corresponds to higher nearby star formation activity, is exposed to a stronger ISRF, saturating at gas densities of $n_{\mathrm{H}}\gtrsim10\,\mathrm{cm}^{-3}$ (see their Fig.~1). Note that \textsc{colibre} uses the ``weak ISRF'' model from \citet{P25}, in which the ISRF is reduced by a factor of 10 compared to their fiducial value (see discussion in Sect.~\ref{sec:discussion_model}). The radiation is shielded by a column density, $N_{\mathrm{sh}}$, which depends on the local Jeans column density (Fig.~3 in \citealp{P25}; see also \citealp{Schaye2001, PS20}). Cosmic ray ionization is largely constant within the ISM (Fig.~2 in \citealp{P25}) and  \textsc{colibre} uses an \ion{H}{I} ionization rate from cosmic rays\footnote{More recent work suggests that the cosmic ray ionization rate may be a factor of $\approx 9$ lower \citep{Obolentseva2024}.} of $\zeta_0 = 2\times10^{-16}\,\mathrm{s}^{-1}$ from \citet{Indriolo2015}. Note that while this work focuses on the ISM in galaxies, the assumptions on radiation fields, shielding columns, and CR rates in \textsc{colibre} are required to be appropriate for all gas throughout the simulation box, including circum- and intergalactic gas.

    The element abundances of H, He, C, N, O, Ne, Mg, Si, S, Ca, and Fe in \textsc{colibre} change through time- and metallicity-dependent stellar mass loss of nearby stars, based on the nucleosynthetic yields compiled by \cite{Correa2026}. The individual species fractions are calculated with the chemical network \textsc{chimes} \citep{Richings2014optthin,Richings2014shielded}. For H and He species, these are evaluated in non-equilibrium through the \textsc{chimes} integration in \textsc{swift}, while the species abundances for the remaining elements (using the same prescriptions for radiation fields, shielding column densities and cosmic ray rates) are pre-tabulated under the assumption of chemical steady-state and ionization equilibrium. Note that \citet{P25} and \citet{Ploeckinger2025_H2} showed that a reduced network of only H and He species results in an H$_{2}$ excess close to the \ion{H}{I}-H$_2$ transition. This is mainly caused by the lack of reactions with oxygen species, which significantly contribute to the destruction of H$_2$. In this work, we focus on the transition from warm to cold \ion{H}{I} gas, which is not affected. 

    The early feedback model \citep{BenitezLlambay2026} controls gas particles tagged as \ion{H}{II} regions. The species fractions of these particles are set to $n_{\mathrm{HII}}/n_{\mathrm{H}} = n_{\mathrm{HeII}}/n_{\mathrm{He}} = 1$ and are stopped from cooling below a temperature of $10^4~\rm K$. After an allotted time (connected to the lifetime of young massive stars), the abundances and temperature are evolved by \textsc{hybrid-chimes} again. 
    Only a small fraction of the gas is in \ion{H}{II} regions at any given time so they have only a small impact on the properties of the ISM reported in this work. 

    Not all atoms remain in the gas phase; some are depleted onto dust grains. In \textsc{colibre}, dust grains of three different chemical compositions (graphites and two kinds of silicates) are seeded by AGB stars and core-collapse supernovae (CCSNe) and the dust mass grows efficiently in the neutral ISM through accretion. In hot gas, dust grains are destroyed by thermal sputtering. Dust grain sizes may increase (through coagulation) or decrease (through grain shattering), which is tracked for two different grain-size bins per dust composition (see \citealp{Trayford2025} for details). The H and He non-equilibrium chemistry network is fully coupled to the live-dust model. The evolving dust properties affect, e.g. the shielding of the far-UV (FUV) radiation from the ISRF and the formation of molecular hydrogen. Most importantly for this work, the PE heating from dust grains depends on the time-variable dust content. 
    
    As mentioned above, the metal species abundances are pre-tabulated and assume a constant dust-to-metal ratio within the ISM. Nevertheless, the metal cooling rates from \textsc{hybrid-chimes} \citep{P25} are connected to both the live-dust model and the non-equilibrium abundances of hydrogen and helium species, since (i) the cooling and heating rates for each element are reduced by the depletion fractions from the live-dust model, and (ii) the metal cooling rates are corrected to account for the non-equilibrium electron fractions from H and He species.

\subsubsection{Galaxy properties}\label{sec:galaxy_properties}

    The \textsc{colibre} simulations are post-processed to locate structures like galaxies and clusters of galaxies. A friends-of-friends (FoF) halo finder is applied first to dark matter particles, with baryonic particles then assigned to their nearest dark matter neighbour. 
    The subhalo finder \textsc{hbt-herons} \citep{ForouharMoreno2025}, which is an updated version of \textsc{hbt+} \citet{Han2012, Han2018}, is applied to the snapshots with the FoF catalogue as input. \textsc{hbt-herons} tracks the hierarchical formation of self-bound objects across simulation outputs. 
    The output of \textsc{hbt-herons} is used by the spherical overdensity and aperture processor \textsc{soap} \citep{McGibbon2025} to calculate a catalogue of (sub-)halo and galaxy properties in different apertures. 

    In our analysis, we utilize the \textsc{soap} catalogue for the galaxy selection and describe the properties used here\footnote{\textsc{soap} is publicly available at \href{https://github.com/SWIFTSIM/SOAP}{\texttt{github.com/SWIFTSIM/SOAP}}.}. For the halo mass, $M_{\mathrm{halo}}$, we use the mass of a spherical overdensity that encloses an average density of 200 times the critical density of the Universe. All other galaxy properties are evaluated for all gravitationally bound particles within a 3D aperture of radius 50 proper kpc (pkpc)\footnote{\citet{Chaikin2025} discuss the dependence of the galaxy stellar mass function on the choice of aperture and find negligible differences for $M_\star \lesssim 10^{11} \mathrm{M}_\odot$, which is the mass range of the galaxies analysed in this work.}, centred on the most bound particle of the halo. A galaxies stellar mass, $M_{\star}$, is the sum of the masses of the star particles and the ISM mass, $M_{\mathrm{ISM}}$, is the sum of the masses of the cool ($T<10^{4.5}\,\mathrm{K}$), dense ($n_{\mathrm{H}}>0.1\,\mathrm{cm}^{-3}$) gas particles. The ISM metallicity, $Z_{\mathrm{ISM}}$, is the average metal mass fraction (including both gas and dust) of this ISM gas. 
    For the specific star formation rate, $\mathrm{sSFR}$, the sum of the instantaneous star formation rate ($\mathrm{SFR}$) of the gas particles is divided by $M_{\star}$. 

    \begin{figure*} 
        \centering
        \includegraphics[width=0.98\linewidth]{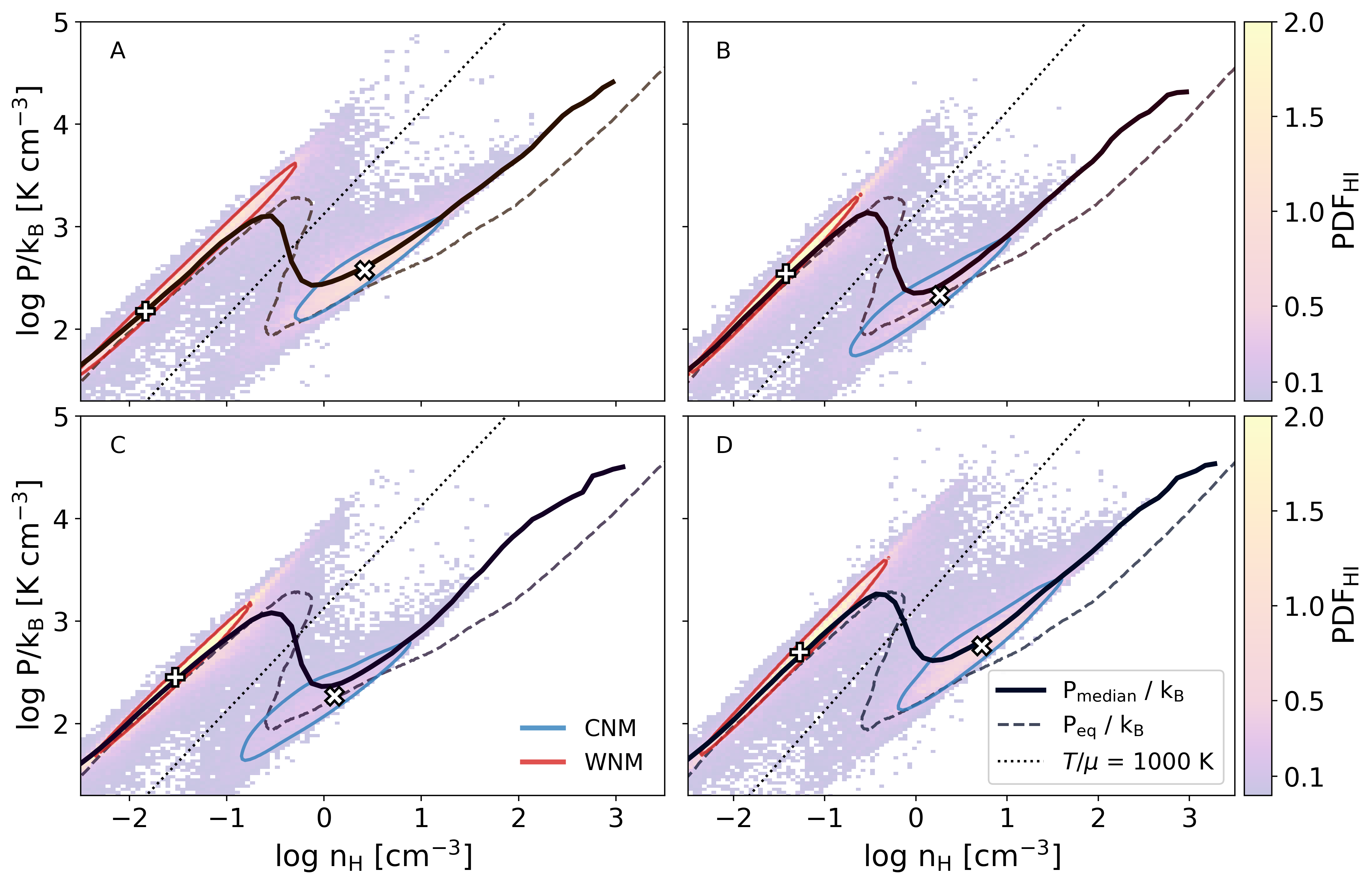}
        \caption{The thermal pressure $P/k_\mathrm{B}$ as a function of hydrogen number density $n_\mathrm{H}$ for four galaxies with solar metallicity in the L025m5 simulation. In each galaxy the gravitationally bound particles within a 3D aperture of radius 50~pkpc are selected.
        The 2D histograms show the probability density function (PDF) weighted by the atomic hydrogen mass $M_\mathrm{HI}$ and overlaid are the median (solid lines) and thermal equilibrium (dashed lines) pressures, and the temperature threshold separating the CNM and WNM (dotted lines). The separate PDFs ($M_\mathrm{HI}$ weighted) for the cold and warm neutral medium are shown as a blue and red contours (containing $\approx 50\%$ \ion{H}{I} mass), respectively. The maxima of these PDFs are marked with plus signs (WNM) and crosses (CNM). 
        }
        \label{fig:MW_hist_med_4} 
    \end{figure*}
    
    \begin{figure*}
        \centering
        \includegraphics[width=0.9\linewidth]{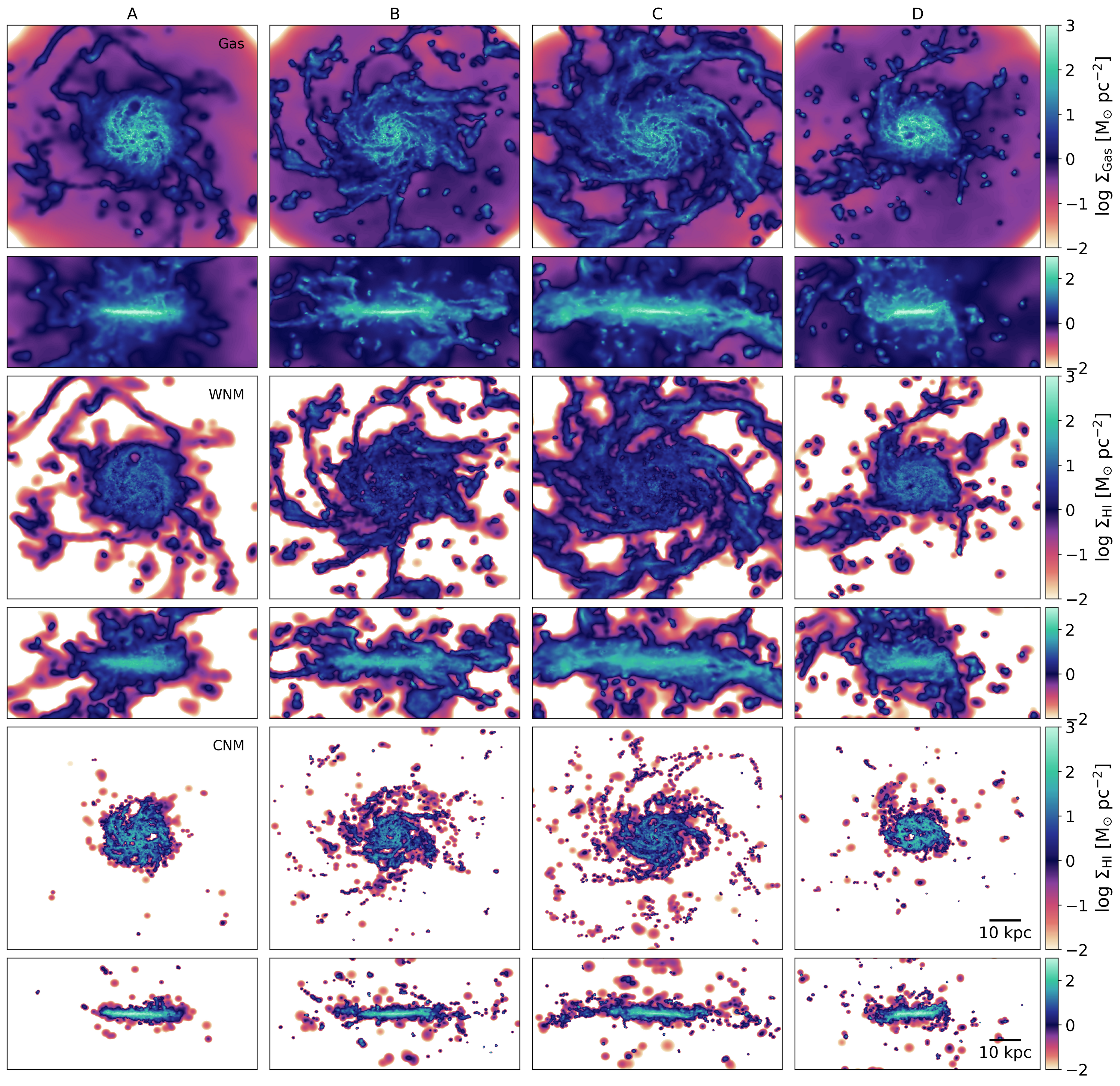}
        \caption{Face-on and edge-on views of the surface mass density of all gas $\Sigma_\mathrm{Gas}$ (first and second row), atomic hydrogen in the WNM $\Sigma_\mathrm{\ion{H}{I}}$ (third and fourth row), and CNM $\Sigma_\mathrm{\ion{H}{I}}$ (fifth and sixth row) for the four \textsc{colibre} galaxies with solar metallicity in the L025m5 simulation are shown in each column. 
        The gas particles are split into cold and warm phase by a temperature threshold of $T/\mu = 1000\, \mathrm{K}$.
        A scale bar is shown in the lower right corner of the figure. Some properties of these galaxies can be found in Table \ref{tab:MW_ex_properties}.}
        \label{fig:MW_surf_dens} 
    \end{figure*}
    
\section{Results}\label{sec:results}

In this section, we present the thermal pressure distribution of the neutral ISM in galaxies within the \textsc{colibre} suite of simulations.
The multiphase structure of the cold and warm neutral media in disk galaxies with approximately solar metallicity is studied in Sect.~\ref{sec:MW_analogue}. 
The metallicity dependence of the thermal pressure distribution and the separate ISM phases is presented in Sect.~\ref{sec:metallicity_dependence}.
Note that the thermal pressures of \textsc{colibre} galaxies are determined based on gas particle data\footnote{See equation~\ref{eq:eq_pressure}, where the equilibrium quantities are replaced by those from the gas particle data.}, where all particles within 50~pkpc of the centre of each galaxy are included.

\subsection{Galaxies with solar-metallicity ISM}
\label{sec:MW_analogue}

    \begin{figure*} 
        \centering
        \includegraphics[width=0.95\linewidth]{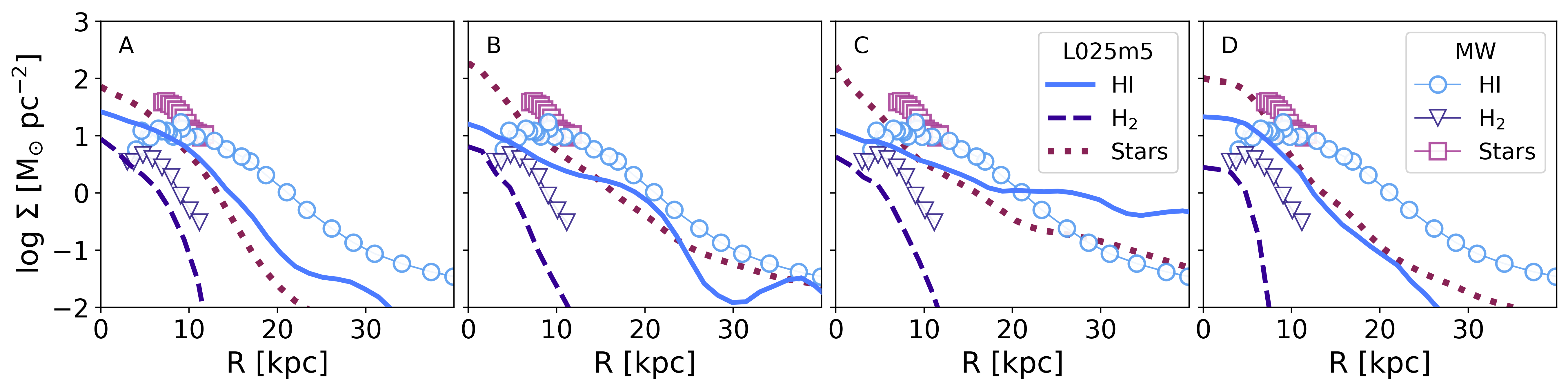}
        \caption{The median mass surface density distribution of stars (dotted line), molecular (dashed line), and atomic (solid line) hydrogen as a function of the radius $R$~[kpc] for four galaxies with approximately solar metallicity in the L025m5 simulation. The radial distribution of stellar mass surface density (\protect\citealp{Xiang2018}; squares), the H$_2$ mass surface density (from Fig.~7 of \protect\citealp{HeyerDame2015}; triangles), and the average surface density of the total \ion{H}{I} disk (\protect\citealp{KalberlaKerp2009}; circles) of the Milky Way are added for reference. 
        }  
        \label{fig:Radial_surf_dens_MW}
    \end{figure*}
    
    \begin{figure} 
        \centering
        \includegraphics[width=0.95\linewidth]{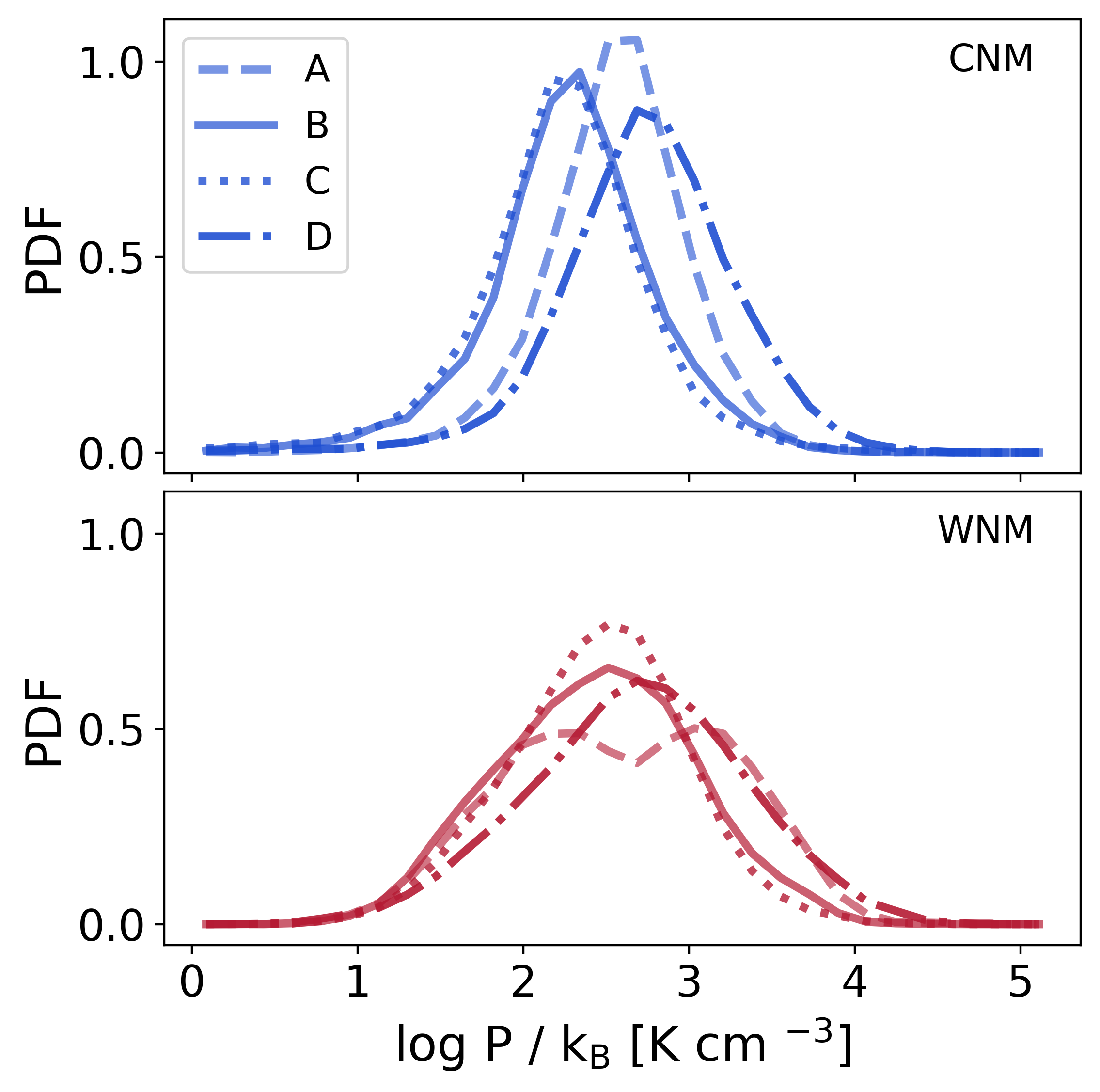}
        \caption{
            The thermal pressure distributions of the cold (top) and warm (bottom) neutral media of the four galaxies with solar metallicity in the L025m5 simulation. The gas particles are split by a temperature threshold of $T/\mu = 1000\, \mathrm{K}$ into a cold and a warm phase. The pressure PDFs are weighted by the \ion{H}{I} mass of each gas particle.} 
        \label{fig:MW_1D_pressure_hist} 
    \end{figure}

    We randomly select four central, star-forming ($\mathrm{sSFR > 10^{-2}\,\mathrm{Gyr}^{-1}}$) galaxies with an average ISM metallicity close to solar ($\log Z_\mathrm{ISM}/\mathrm{Z}_{\odot} = 0\,\pm \,0.1 \,\mathrm{dex}$) and a well-resolved ISM ($\log M_{\mathrm{ISM}}\,[\mathrm{M}_{\odot}]\ge 9.5$) from the 24\footnote{The figures from the analysis of all 24 galaxies are available at \texttt{To be added upon publication}}. objects in L025m5\footnote{At the time of writing, L025m5 is the largest available volume at m5 resolution for $z=0$. The L050m5 and L100m5 \textsc{colibre} simulations are currently being evolved.} 
    at $z=0$ that fulfil these criteria. We use this small sample to exemplify the thermal pressure distributions of the multiphase ISM in individual objects, before we investigate the full galaxy population in Sect.~\ref{sec:metallicity_dependence}. 
    
    The main properties of the selected galaxies are listed in Table~\ref{tab:MW_ex_properties} and the values for the MW are included for reference. We verified that the \ion{H}{I} mass-weighted average metallicities are very similar to the ISM metallicities.
    Because of the small volume and random phases of the initial conditions (i.e. not constrained to reproduce the MW), this sample does not consist of direct MW twins, but is used to illustrate the diversity of thermal pressure distributions within a narrow range of gas metallicities.
    
    Fig.~\ref{fig:MW_hist_med_4} shows the distribution of pressures and densities of \ion{H}{I} gas in the selected galaxies. The 2D histogram in the background shows the probability density function (PDF) of all gas particles, weighted by their \ion{H}{I} masses. The warm and cold phases are separated by the dotted line at $T/\mu=1000\,\mathrm{K}$ and the respective contours indicate the pressure and density ranges containing approximately $50\%$ of the \ion{H}{I} mass of the CNM (blue) and the WNM (red). 
    We see from the 2D histogram of the PDF that the region between the WNM and CNM contours, the unstable neutral medium, is sparsely populated and our results are therefore not sensitive to the exact value of the temperature that separates the phases. The peaks of the phase-separated PDFs\footnote{We use Kernel Density Estimation (KDE) with Gaussian functions to obtain the separate PDFs weighted by the \ion{H}{I} masses in each phase. The full 2D distribution is not determined using Gaussian KDE, since they tend to over-smooth bimodal distributions. The blue (CNM) and red (WNM) contours and peaks of each PDF in Fig.~\ref{fig:MW_hist_med_4} are derived from the KDE.} are highlighted with plus signs and crosses for warm and cold gas, respectively. 

    The black solid lines in Fig.~\ref{fig:MW_hist_med_4} indicate the median pressure as a function of gas density. These reveal the classical S-shape from thermal equilibrium calculations, but it is unable to capture the complete pressure range of the multiphase gas compared to the full gas distribution, represented by the 2D histogram and the contours. 
    Finally, we show the thermal equilibrium pressures for solar metallicity gas (dashed line) for reference. While the equilibrium pressures assume thermal equilibrium, a constant dust-to-metal ratio ($\mathcal{DTM} = 0.49$), constant depletion fractions, solar abundance ratios, steady-state chemistry and ionization equilibrium (see Appendix~\ref{sec:thermalequilibrium} for details), the ISM phases of the simulated galaxies loosely follow the equilibrium pressure $P_{\mathrm{eq}}$. This allows us to use equilibrium functions to interpret the relevant processes in Appendix~\ref{sec:appendix:metdependenceexplanation}. 

    Fig.~\ref{fig:MW_surf_dens} shows the ISM morphologies of the selected galaxies, both in face-on (larger panels) and edge-on (smaller panels) projections, displaying the mass surface densities of all gas (top), \ion{H}{I} in the WNM with $T/\mu>1000\,\mathrm{K}$ (middle), and \ion{H}{I} of the CNM with $T/\mu\leq 1000\,\mathrm{K}$ (bottom). 
   
    While these galaxies exhibit a disk-like structure, their gas distributions range from compact (A, D) to extended (B, C). Furthermore, the gas surface density of Galaxy~D reveal a central \ion{H}{I} hole indicative of recent feedback events, consistent with the high star formation rate of this object (Table~\ref{tab:MW_ex_properties}). For reference, we show the stellar, \ion{H}{I}, and H$_{2}$ radial surface density profiles for these galaxies against the measured radial profiles for the MW \citep{Xiang2018, HeyerDame2015, KalberlaKerp2009} in Fig.~\ref{fig:Radial_surf_dens_MW}. 
    This quantitatively confirms the impression that galaxies A and D have more compact \ion{H}{I} disks than the MW. 

    The 1D thermal pressure PDFs of cold (top) and warm (bottom) gas of the four objects are shown in Fig.~\ref{fig:MW_1D_pressure_hist}. It shows that, for these galaxies, the WNM and CNM coexist in a specific range of thermal pressures, mainly between $\log P_{\mathrm{min}}/k_{\mathrm{B}}\,[\mathrm{K\,cm}^{-3}]\approx 2$ and $\log P_{\mathrm{max}}/k_{\mathrm{B}}\,[\mathrm{K\,cm}^{-3}]\approx 3$. Higher WNM pressures (up to $\log P/k_{\mathrm{B}}\,[\mathrm{K\,cm}^{-3}]\approx 4$) are common, whereas CNM pressures of $\log P/k_{\mathrm{B}}\,[\mathrm{K\,cm}^{-3}]\gtrsim 3.5$ are mainly seen in object~D (bottom right), a galaxy with a relatively high SFR of $4.3\,\mathrm{M}_{\odot}\,\mathrm{yr}^{-1}$ and H$_2$ mass of $3.36\times 10^{9}\, \rm M_\odot$.  
    We find a negative metallicity gradient and a decrease in the pressure of both phases with increasing galactocentric radius in these \textsc{colibre} galaxies (see Appendix~\ref{sec:appendix:radial} for details). 
    
    We conclude that the four \textsc{colibre} galaxies selected within a narrow range in average ISM gas metallicity ($\log Z_{\mathrm{ISM}}/\mathrm{Z}_{\odot} = 0 \pm 0.1$) display a consistent picture of their multiphase structure, despite their different gas morphologies and disk sizes. Indeed, each galaxy exhibits well-separated warm and cold neutral gas phases that coexist within a certain range of thermal pressures, generally between $\log P/k_{\mathrm{B}} \,[\mathrm{K\,cm}^{-3}]= 2$ and $\log P/k_{\mathrm{B}} \,[\mathrm{K\,cm}^{-3}]= 3$, but with tails extending to both higher and lower pressures. The particular choice of galaxies has little impact on the conclusions. All 24 galaxies that satisfy the selection criteria show qualitatively similar pressure-density distributions.

\subsection{Metallicity dependence}\label{sec:metallicity_dependence}
    
    \begin{figure}
        \centering
        \includegraphics[width=\linewidth]{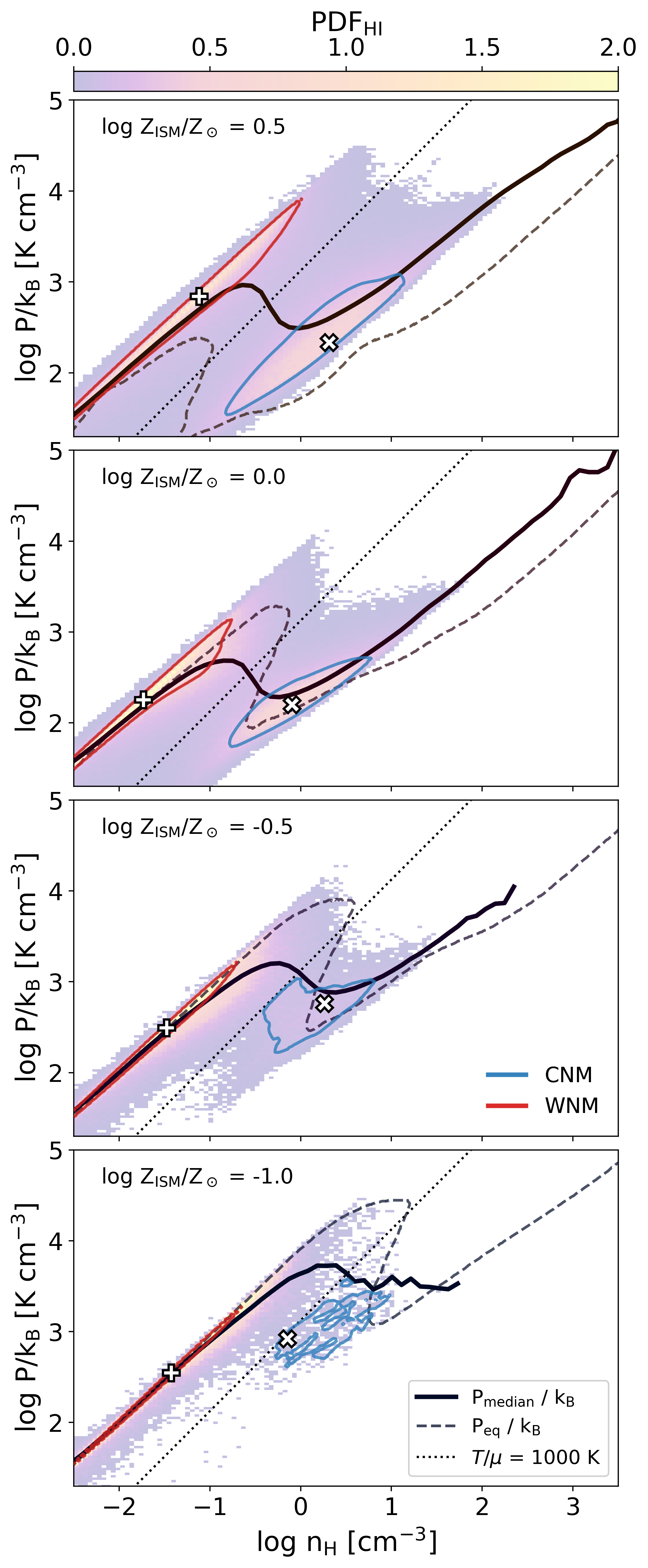} 
        \caption{The thermal pressure $P/k_\mathrm{B}$ as a function of hydrogen number density $n_\mathrm{H}$ for galaxies in the L200m6 simulation with average ISM metallicity, $\log Z_\mathrm{ISM}/Z_\odot = [-1,\,-0.5,\,0,\, 0.5] \,\pm\, 0.1$, from top to bottom. 
        The components of each panel are the same as in Fig.~\ref{fig:MW_hist_med_4} but for the combination of all particles in the 500 selected galaxies in each metallicity bin. 
        }
        \label{fig:all_Z_hist_med_contour} 
    \end{figure}

    \begin{figure}
        \centering
        \includegraphics[width=0.98\linewidth]{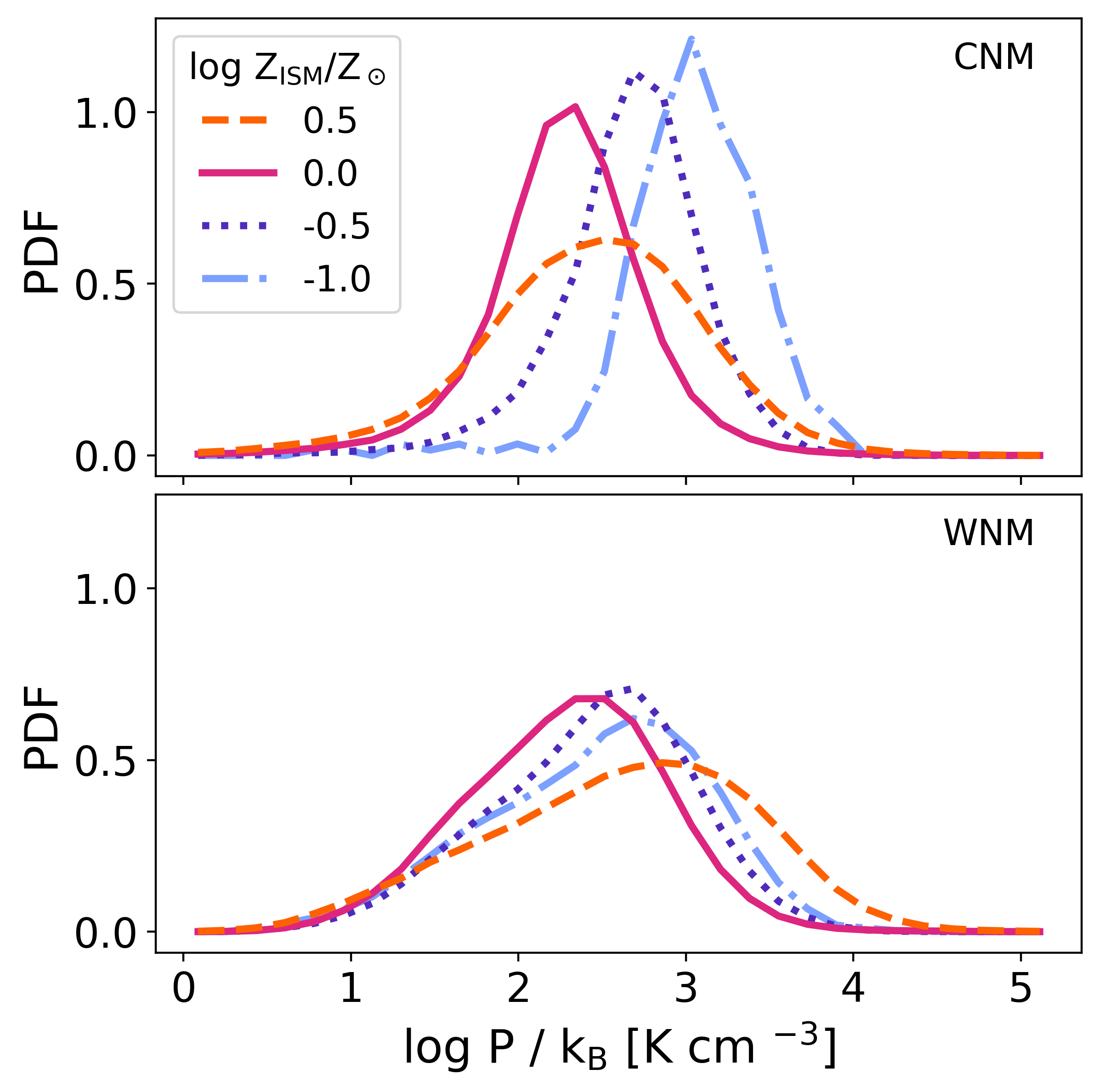}
        \caption{The thermal pressure distributions for all metallicity bins. The gas is split into the cold (top panel) and warm (bottom panel) neutral medium by temperature $T/\mu = 1000\, \mathrm{K}$. The pressure PDFs are weighted by the particles \ion{H}{I} mass fraction for the respective phases. } 
        \label{fig:Pk_1D_hists} 
    \end{figure}

In this section, we focus on the multiphase ISM of large galaxy populations. We use the largest simulated cosmological volume with a side length of 200~cMpc at the intermediate (m6) resolution. Compared to the analysis of the high-resolution (m5) simulation in Sect.~\ref{sec:MW_analogue}, the mass resolution is a factor of 8 lower. In Sect.~\ref{sec:resolution}, we find that the thermal pressure of the neutral phases does not converge with the numerical resolution and that a smaller simulation (L025m6) reproduces the general trend of the larger simulation (L200m6).

We select central, star-forming ($\mathrm{sSFR > 10^{-2}\,\mathrm{Gyr}^{-1}}$) galaxies within narrow bins of ISM gas metallicity to study the metallicity dependence of the multiphase ISM in \textsc{colibre}. The individual bins are defined as $\log Z_{\mathrm{ISM}}/\mathrm{Z}_{\odot} = [-1,\,-0.5,\,0,\, 0.5] \,\pm\, 0.1$. The total number of galaxies within each bin in the L200m6 simulation ranges from 650 to 42,000, and we select a random subsample of 500 objects per metallicity bin.

\textsc{colibre}, and in particular the L200m6 simulation, matches the observed relation between stellar mass and gas-phase metallicity (from oxygen abundance) from \citet{Curti2020} for star-forming galaxies \citep{Schaye2025colibre}\footnote{Different observational studies find different tracks for the mass–metallicity relation and the \textsc{colibre} simulations at different resolutions cover these tracks but they are not well converged with numerical resolution at $z=0$ (see Fig.~20 in \citealp{Schaye2025colibre}).}. Similarly, the stellar mass-stellar metallicity relation of L200m6 matches observations across several orders of magnitude in stellar mass, with deviations only for the most massive galaxies with $M_{\star}\gtrsim10^{11}\,\mathrm{M}_{\odot}$, which are typically quenched \citep{Schaye2025colibre}. 
Note that the redshift $z=0$ median gas-phase oxygen abundance as a function of stellar mass for star-forming galaxies in L200m6 saturates at $12 +\log(\mathrm{O}/\mathrm{H})\approx 8.8$ ($\log Z/\mathrm{Z}_{\odot}\approx 0.1$ for solar abundance ratios) and only increases further to $12 +\log(\mathrm{O}/\mathrm{H})\approx 9$ for the most massive galaxies (see Fig.~20 in \citealp{Schaye2025colibre}). Galaxies in the highest metallicity bin, $\log Z_\mathrm{ISM}/ \mathrm{Z}_{\odot} = 0.5$ ($12 +\log(\mathrm{O}/\mathrm{H})\approx 9.11$), lie above the stellar mass - gas metallicity relation, and are therefore biased towards objects with a high gas-phase metallicity for their stellar mass. We find that, as predicted by the fundamental relation between stellar mass, SFR, and gas-phase metallicity of observed galaxies \citep{Mannucci2010}, the mean SFR of this galaxy sample is biased towards low values for their stellar mass. 

At lower gas metallicities, the rates of both important cooling (metal-line cooling) as well as heating (PE heating) processes decrease. The metallicity dependence of the pressure of the multiphase ISM is therefore not trivial. 
For example, using a theoretical model, \citet{Bialy2019} found that the allowed pressure range of equilibrium pressures for multiphase gas first decreases from $Z_{\mathrm{ISM}} = \mathrm{Z}_{\odot}$ to $Z_{\mathrm{ISM}} = 0.1\,\mathrm{Z}_{\odot}$ but increases beyond the solar metallicity values for metallicities of $Z_{\mathrm{ISM}} \leq 0.01\,\mathrm{Z}_{\odot}$. They assumed for each of their metallicity models the same radiation field, cosmic ray rates, and a dust-to-gas ratio that depends on metallicity via a broken power-law. 
In this work each metallicity bin represents distinct galaxies, exhibiting different dust properties, ISRF strengths, and CR rates depending on temperature and density of the gas. 

Fig.~\ref{fig:all_Z_hist_med_contour} shows the collective pressure-density distributions of all \ion{H}{I} gas within the selected galaxies with the highest ISM metallicity ($\log Z_{\mathrm{ISM}}/\mathrm{Z}_{\odot} = 0.5$, top panel) to those with the lowest ISM metallicity ($\log Z_{\mathrm{ISM}}/\mathrm{Z}_{\odot} = -1$, bottom panel). We do not include lower metallicities because in those cases the transition to the cold gas phase shifts to higher densities that we do not resolve in our simulations. In lower metallicity objects, star formation also occurs in the warm gas phase; the WNM - CNM transition is therefore less reliable. Furthermore, in the L200m6 simulation at $z=0$, the majority of galaxies with $\log Z_{\mathrm{ISM}}/\mathrm{Z}_{\odot} = -1$ have ISM masses $M_{\mathrm{ISM}}\le 10^8\,\mathrm{M}_{\odot}$ (see Appendix~\ref{sec:appendix:metallicity_dependence}), which is represented by fewer than 100 gas particles. 

We characterize the different phases, as well as their transition, in Fig.~\ref{fig:all_Z_hist_med_contour} as previously done in Fig.~\ref{fig:MW_hist_med_4}. The fraction of neutral gas that is in the cold phase decreases with decreasing metallicity. 
For the galaxies with $\log Z_{\mathrm{ISM}}/\mathrm{Z}_{\odot} \ge -0.5$, the pressure range of multiphase ISM gas remains largely between $\log P_{\mathrm{min}}/k_{\rm B}\,[\mathrm{K\,cm}^{-3}] = 1.5$ and $\log P_{\mathrm{min}}/k_{\rm B}\,[\mathrm{K\,cm}^{-3}] = 3.0$.
Only for galaxies with the lowest metallicities ($\log Z/\mathrm{Z}_{\odot} = -1$) does the cold phase shift to higher pressures, though very little cold gas is present in their ISM. The metallicity-dependent thermal pressure distributions of the cold (top panel) and warm (bottom panel) media are shown in Fig.~\ref{fig:Pk_1D_hists}. The peak of the pressure PDFs shifts to higher pressures for lower metallicities for $\log Z/\mathrm{Z}_{\odot} \le 0$. The thermal pressures in the highest metallicity bin do not follow this trend.

Using equilibrium functions (Appendix~\ref{sec:thermalequilibrium}), we analyse the origin of the metallicity dependence of the multiphase ISM in \textsc{colibre}. Lower gas metallicities lead to a decrease in C and H$_2$ related cooling rates and for $\log Z/Z_\odot \leq 0$ the dust related heating rates decrease for intermediate temperatures and a gas density of $\log n_{\rm H}\, [{\rm cm}^{-3}] = 0$ (see Appendix~\ref{sec:appendix:metdependenceexplanation} for details). The total heating and cooling rates are comparable across large temperature ranges. This makes the equilibrium temperature, $T_{\mathrm{eq}}$, where heating and cooling rates are balanced, very sensitive to minor changes in the model, particularly around solar metallicity. 
The median pressures of the simulated ISM deviate from the equilibrium pressures (Fig.~\ref{fig:all_Z_hist_med_contour}), especially in the multiphase range, since $T_\mathrm{eq}$ and $P_\mathrm{eq}$ are sensitive to model variations. 

We conclude that \textsc{colibre} reproduces the complexity of the multiphase ISM through the non-equilibrium species abundances, non-solar relative abundances, mechanical heating and cooling, and especially the live-dust model in \textsc{colibre}, which affect both heating and cooling rates in a non-linear way.

\section{Discussion}\label{sec:discussion}

\subsection{The COLIBRE ISM model}\label{sec:discussion_model}

In this section, we discuss the thermal pressure distributions of the ISM in \textsc{colibre} galaxies within the context of the \textsc{colibre} ISM model. In particular, we discuss the influence of the interstellar radiation field on the results (Sect.~\ref{sec:disc:weakISRF}) and demonstrate the impact of the live-dust model on the $\mathcal{DTG}$ ratio of the WNM (Sect.~\ref{sec:disc:dust}). The influence of resolution (L025m5, L025m6, L025m7) and box size (L025m6, L200m6) is discussed in Sect.~\ref{sec:resolution}. 

\subsubsection{The interstellar radiation field in COLIBRE}\label{sec:disc:weakISRF}

\begin{figure}
    \centering
    \includegraphics[width=0.98\linewidth]{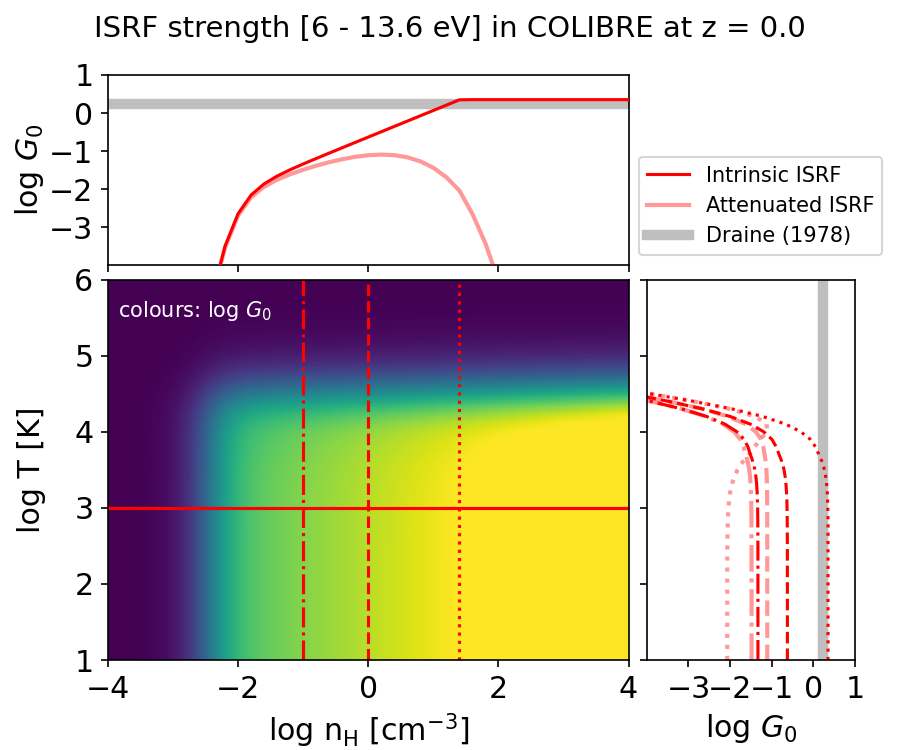}
    \caption{The intrinsic ISRF assumed in \textsc{colibre}. The brighter colours in the large panel represent a stronger intrinsic radiation field (here shown for the energy range [$6-13.6\,\mathrm{eV}$] in \citet{Habing1968} units, $G_0$). 
    The small panels show the intrinsic ISRF strengths for constant temperature (top; see matching horizontal line in large panel) and density (right; see matching vertical lines in large panel). Gas with low temperatures ($T\lesssim 10^4 \,\mathrm{K}$) and high densities is exposed to the strongest intrinsic ISRF, reflecting the higher local SFRs. The intrinsic ISRF is an input into \textsc{hybrid-chimes} which processes the radiation through a shielding column (see Sect.~\ref{sec:hybrid-chimes} and \citealp{P25} for details). The resulting attenuated ISRF is used to calculate the net cooling rates of the gas. The lighter lines in the small panels show the attenuated ISRF for the dust-to-gas ratio assumed in the chemical equilibrium tables for reference. In \textsc{colibre}, the attenuation for each density and temperature may vary because the dust content within the shielding column is determined by the live-dust model. The intrinsic ISRF in \textsc{COLIBRE} matches the canonical value of $G_0=1.7$ from \citet{Draine1978} (thick, grey horizontal and vertical lines in the top and right panels, respectively) at high densities ($\log n_{\mathrm{H}}\, [\mathrm{cm}^{-3}]\ge 1.4$) and is $\approx 1\,\mathrm{dex}$ lower at densities typical for the WNM ($\log n_{\mathrm{H}}\, [\mathrm{cm}^{-3}]\approx -0.5 $).
    } 
    \label{fig:G0}
\end{figure}

The intensity of the ISRF is typically normalized to the solar-neighbourhood value from either \citet{Habing1968}, \citet{Draine1978}, or \citet{Mathis1983}, which are all within a factor of 1.7 of each other. 
The \textsc{colibre} flagship simulations use the ``weak ISRF'' model\footnote{\cite{Schaye2025colibre} note that a stronger ISRF, such as the fiducial normalization from \citet{P25}, results in a higher fraction of star formation in warm ($T \sim 10^4\,\mathrm{K}$) gas for dwarf galaxies ($M_\star \sim 10^9\,\mathrm{M}_\odot$), due to the limited resolution, compared to the fiducial \textsc{colibre} model. Another motivation for the ``weak ISRF'' model is the separate implementation of \ion{H}{II} regions.} from \citet{P25}, which at typical WNM densities gives ISRF strengths an order of magnitude lower than the solar-neighbourhood value. As explained in Sect.~\ref{sec:hybrid-chimes} (see \citealp{P25} for details), the radiation field in \textsc{colibre} depends on gas density and temperature. In Fig.~\ref{fig:G0}, we show the radiation field strength of the ISRF assumed in \textsc{colibre} as the integrated energy flux, $u$, in the energy range between 6 and 13.6~eV in \citet{Habing1968} units, defined as $G_0 = u /[1.6\times10^{-3}\, \mathrm{erg}\ \mathrm{cm}^{-1}\,\mathrm{s}^{-1}]$. 
Motivated by increased star formation activity in denser gas, $G_0$ increases with density (proportional to the Jeans column density, motivated by the KS law; \citealp{Schmidt1959, Kennicutt1989}) in neutral gas and saturates at $\log n_{\mathrm{H}}\,[\mathrm{cm}^{-3}]\approx 1.5$. The ISRF in \textsc{colibre} slightly exceeds the canonical value of $G_0 = 1.7$ (\citealp{Draine1978}; thick solid lines in the small panels of Fig.~\ref{fig:G0}) at $\log n_{\mathrm{H}}\,[\mathrm{cm}^{-3}]\geq 1.4$ (dotted line), but it is 0.6~dex and 1.3~dex weaker at densities of $\log n_{\mathrm{H}}\,[\mathrm{cm}^{-3}]= 0$ (dashed line) and $\log n_{\mathrm{H}}\,[\mathrm{cm}^{-3}]= -1$ (dash-dotted line), respectively. These are typical densities for the WNM - CNM transition in solar-metallicity gas.
The ISRF strength in \textsc{colibre} applies to large density ranges, not only densities present in the ISM of galaxies, which explains the low values for $G_0$ at densities of $\log n_{\mathrm{H}}\,[\mathrm{cm}^{-3}]\lesssim -2$.

We have run a small \textsc{colibre} box (L25m6) with a ten times stronger ISRF and find that the WNM - CNM transition is only slightly shifted towards higher pressures, compared to the standard \textsc{colibre} ISRF (to be discussed in Sect.~\ref{sec:disc:theory}). This shift is smaller than expected from the equilibrium model (see, e.g. the second panel of Fig.~10 in \citealp{P25}) due to the non-equilibrium chemistry and the live-dust model in \textsc{colibre}.
This highlights the limitations of thermal equilibrium models to estimate the physical properties of the stable phases in the ISM (see \citealp{P25} for a similar conclusion).

\subsubsection{The dust-to-gas ratio}\label{sec:disc:dust}

\begin{figure}
    \centering
    \includegraphics[width=0.98\linewidth]{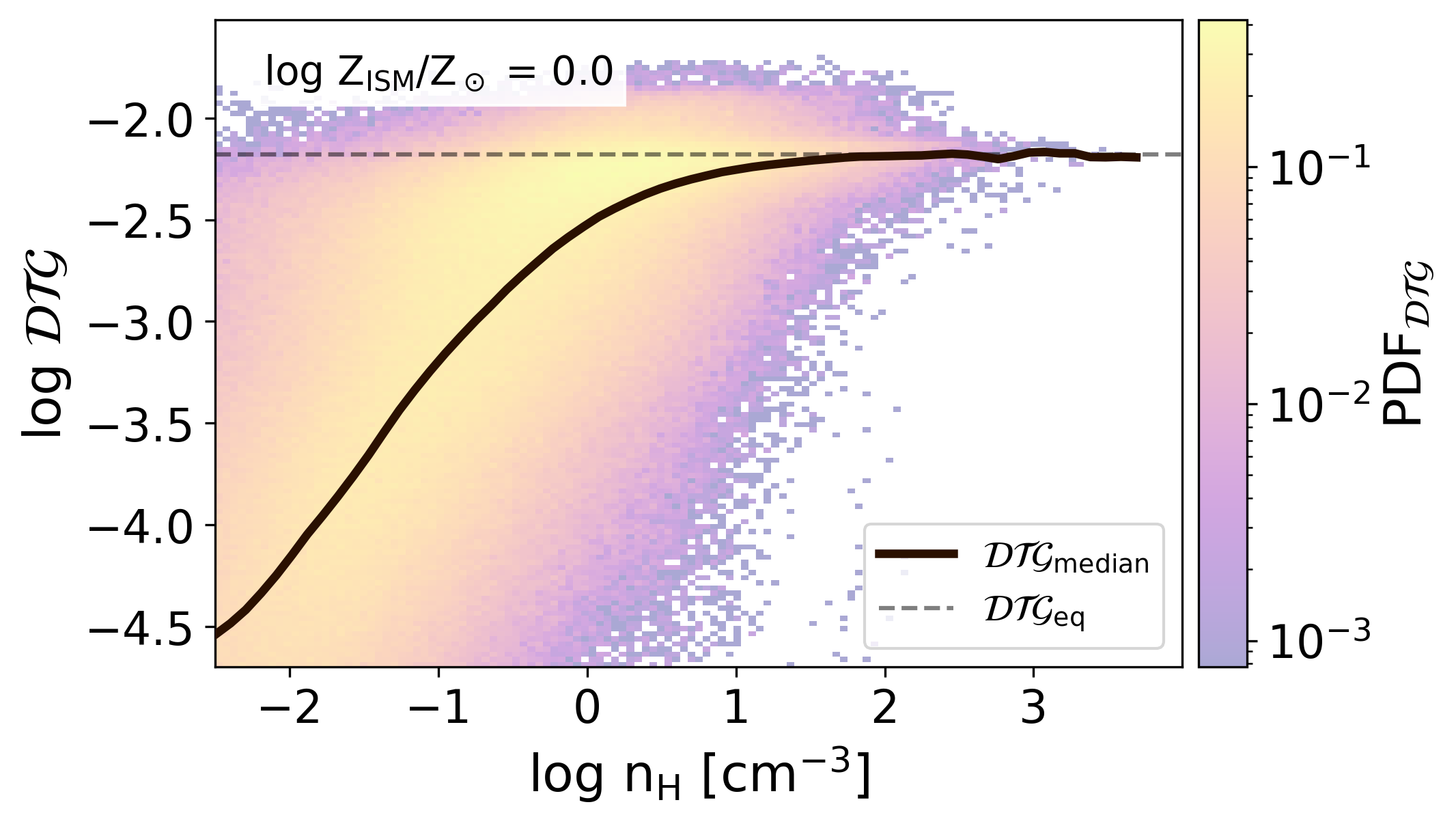}
    \caption{The dust-to-gas mass ratio $\mathcal{DTG}$ (solid line) as a function of hydrogen number density $n_\mathrm{H}$ of particles in 500 galaxies in the L200m6 simulation with average ISM metallicity of $\log Z_\mathrm{ISM}/Z_\odot = 0.0 \,\pm\, 0.1$.
    We show the probability density weighted by the $\mathcal{DTG}$ (histogram), the median (solid line), and the equilibrium value $\mathcal{DTG}_\mathrm{eq} = 6.6\times 10^{-3}$ (dashed line) from the cooling tables. 
    }
    \label{fig:DTZ_solar_Z_bin}
\end{figure}

The live-dust model \citep{Trayford2025} coupled to the hybrid cooling model \citep{P25} in \textsc{colibre} influences thermal processes through depletion, absorption of FUV radiation (leading to PE heating), and molecule formation. 
We discuss the impact of dust on the thermal processes in the multiphase ISM through the dust-to-gas mass ratio $\mathcal{DTG}$ as a function of gas density (Fig.~\ref{fig:DTZ_solar_Z_bin}) for gas particles in the solar-metallicity galaxy sample from the L200m6 simulation. For each galaxy we include all bound gas particles within 50~pkpc of the centre.
This is compared to the dust-to-gas ratio at solar metallicity used in the equilibrium cooling model, which saturates to $\mathcal{DTG}_\mathrm{ISM} = 6.6\times 10^{-3}$, from the grain model in \textsc{Cloudy} \citep{Ferland1998, Chatzikos2023}, at the densities and temperatures of the neutral phases \citep{P25}. 
At lower densities ($\log n_\mathrm{H}\, [\mathrm{cm}^{-3}] \lesssim -1$) the median $\log \mathcal{DTG}_\mathrm{median}$ (solid line) from the simulation is very low ($\lesssim -3$). It increases at densities where the CNM and WNM coexist (see the second panel of Fig.~\ref{fig:all_Z_hist_med_contour}) and converges to the equilibrium value (dashed line) at densities of $\log n_\mathrm{H}\, [\mathrm{cm}^{-3}] \gtrsim 1$. 

In the live-dust model, dust accretion is most efficient in cool, dense gas \citep{Trayford2025}. Strong diffusion processes are therefore necessary to redistribute dust grains into lower density gas, which was not calibrated in \textsc{colibre}. In the equilibrium model, the $\mathcal{DTG}$ ratio is typically assumed to be constant across all ISM densities, while here it is an emerging quantity. 

The $\mathcal{DTG}$ (and therefore also the $\mathcal{DTM}$ at $\log n_\mathrm{H}\, [\mathrm{cm}^{-3}] = 0$) from the simulation is on average lower than in the equilibrium calculations, which impacts the heating and cooling rates. 
While a low dust abundance affects several individual cooling (e.g. by increasing metal line cooling and decreasing H$_2$ cooling) and heating processes (e.g. by lowering shielding and therefore higher photo-heating rates), it mostly impacts the PE heating rate, which is the dominant heating mechanism at this density (see also Appendix~\ref{sec:appendix:metdependenceexplanation}). 

Even though the total dust abundance decreases toward lower densities, the small-to-large grain ratio tends to be higher at lower densities (see Fig.~3 in \citealp{Trayford2025}). Small grains are a factor of 10 smaller in radius, and therefore contribute 10 times larger surface area per unit dust mass. The live-dust model connects cooling and heating processes to dust grains by their surface-to-volume ratio for a given dust mass and material density \citep{Trayford2025}. 
So while the PE heating rates are proportional to $\mathcal{DTG}$, the rates have a secondary dependence on the grain size. 

The lower $\mathcal{DTG}$ at the densities of the WNM - CNM transition in \textsc{colibre} galaxies contributes to the low thermal pressures, since small changes in the dust abundance significantly affect the equilibrium temperature (as described in Appendix~\ref{sec:appendix:metdependenceexplanation}). The impact of changes in the small-to-large grain ratio on the heating and cooling rates needs further investigation.

    \begin{figure} 
        \centering
        \includegraphics[width=\linewidth]{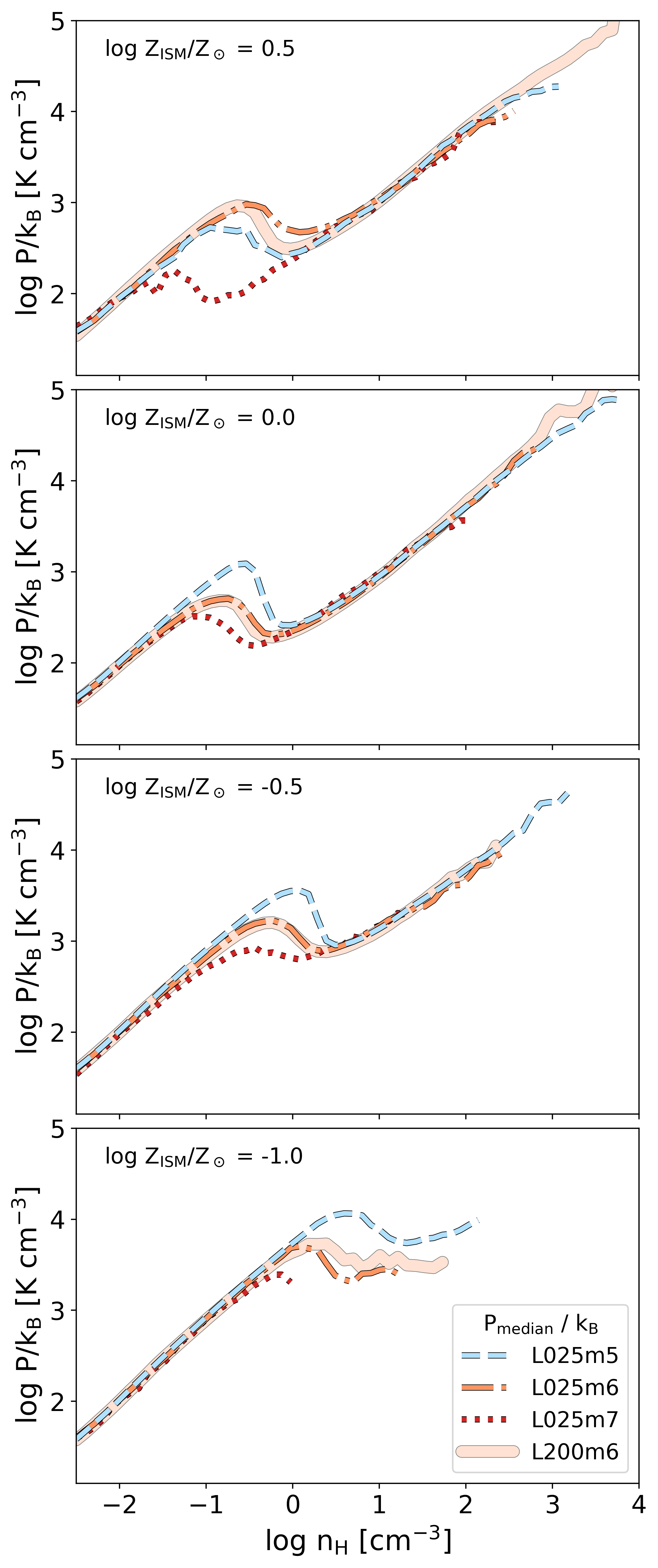}
        \caption{The median thermal pressure $P_\mathrm{median}/k_\mathrm{B}$ as a function of hydrogen number density $n_\mathrm{H}$ for four metallicity bins, $\log Z_\mathrm{ISM}/Z_\odot = [0.5,\,0.0,\,-0.5,\, -1.0] \,\pm\, 0.1$, from top to bottom. The median of all gas particles of 500 galaxies in the L200m6 simulation are depicted as thick transparent lines and the thin dashed, dash-dotted, and dotted lines give the result of galaxies in the L025m7, L025m6, and L025m5 simulations$^{\ref{footnote:Ngalaxies}}$, respectively.
        }
        \label{fig:Compare_res_L25_Z_med} 
    \end{figure}

\subsubsection{Convergence with resolution and box size}\label{sec:resolution}

    The \textsc{colibre} galaxy formation model has been calibrated at three resolution levels (from low to high resolution: m7, m6, and m5). Each resolution level differs by a factor of eight in the initial mass per particle and a factor of two in the constant gravitational force softening length, $\epsilon$ (see Table~\ref{tab:colibre_volumes}). This means that, for each resolution level, gravity is softened at gas densities that are typical for the WNM - CNM transition (in \textsc{colibre}, gravity is softened for  $\log n_{\mathrm{H}}\,[\mathrm{cm}^{-3}]\gtrsim -0.7$ at $z=0$, see Appendix~A in \citealp{Schaye2025colibre} for details). 
    Gravitational processes at these densities are not well resolved and gravitational instabilities are numerically suppressed. For example, \citet{Ploeckinger2024} demonstrated that gravitational instabilities in Lagrangian simulations follow the softened Jeans instability criteria, rather than the standard Newtonian Jeans instability and the typical timescale for gravitational collapse is the longer softened free-fall time. 
    Furthermore, \citet{Benitez2018} showed that the vertical scale-height of self-gravitating disks numerically increases with increasing gravitational force softening length. 
    
    We therefore test the impact of the numerical resolution on the thermal pressure distribution of the multiphase ISM by comparing the galaxy samples\footnote{The number of galaxies in each metallicity bin at each resolution is listed in Appendix~\ref{sec:appendix:Ngalaxies}.\label{footnote:Ngalaxies}} at different metallicities in the L025m7 (red dotted line), L025m6 (orange dot-dashed line), and L025m5 (blue dashed line) simulation in Fig.~\ref{fig:Compare_res_L25_Z_med}. 
    The gas metallicity at redshift $z=0$ in \textsc{colibre} is not well converged (e.g. Fig.~3 in \citealp{Sharda2026}), where simulations at m7 resolution deviate most at high stellar mass and metallicity. By selecting galaxies at the same metallicity, we have galaxies with systematically higher stellar mass at higher resolution. We verified that this has little impact on the conclusions drawn from Fig.~\ref{fig:Compare_res_L25_Z_med}.
    
    Comparing the median thermal pressures of ISM-metallicity-selected galaxies (as in Sect.~\ref{sec:metallicity_dependence}) across the different resolution levels reveals that the thermal pressure distributions are not converged with the numerical resolution. In addition, an improved mass resolution (e.g. from m6 to m5) increases the WNM and CNM pressures by a factor comparable to that resulting from a ten-fold increase in the ISRF strength (see Sect.~\ref{sec:disc:theory}). The non-monotonic behaviour with resolution at the highest metallicity is caused by the small number of galaxies (see Table~\ref{tab:Ngalaxies_resolution}) with this metallicity in a smaller volume (L025). 

    For completeness, the effect of sample size in our analysis is shown by comparing to the galaxy samples in the L200m6 simulation.
    Comparing the median pressures (Fig.~\ref{fig:Compare_res_L25_Z_med}) of galaxies in the L200m6 (thick solid lines) and L025m6 (dash-dotted lines) simulation, shows that a smaller volume containing very few galaxies ($\gtrsim 10$) still reproduces the general trend of the large volume with 500 galaxies. 
    Deviations from the L200m6 sample can be seen in the highest-metallicity bin, where L025m6 hosts only three galaxies.
    We discuss a possible connection of the resolution-dependence of the thermal pressure to the resolution-dependence of the mid-plane pressure in Appendix~\ref{sec:appendix:resolution}. 
    We find that the mid-plane pressures are generally lower in lower resolution simulations and this non-convergence of the mid-plane pressure may contribute to the lack of convergence in the thermal pressures.
    
    We conclude that the thermal pressure range where the warm and cold neutral phase can coexist is not converged with numerical resolution. On the other hand, a smaller sample ($\gtrsim 10$ galaxies) can reproduce the thermal pressure distribution of our larger sample (500 galaxies).

\subsection{Comparison with previous studies}\label{sec:discussion_lit}

We compare the thermal pressures of both the CNM and WNM of \textsc{colibre} galaxies with a mean ISM metallicity of $\log Z_{\mathrm{ISM}} = \log \mathrm{Z}_{\odot} \pm 0.1$ at $z=0$ to values from the literature. In Sect.~\ref{sec:disc:theory}, we discuss theoretical thermal equilibrium models from \citet{Wolfire1995, Wolfire2003}, \citet{Bialy2019}, and \citet{P25}, while in Sect.~\ref{sec:disc:obs}, we focus on measurements from observations in roughly solar metallicity environments \citep{JT11, Gerin2015, Herrera2017}. A comparison to lower metallicity environments \citep{Pineda2017, Kosenko2024} is discussed in Appendix~\ref{sec:disc:obs_lowZ}.

\begin{figure}
    \centering
    \includegraphics[width=0.98\linewidth]{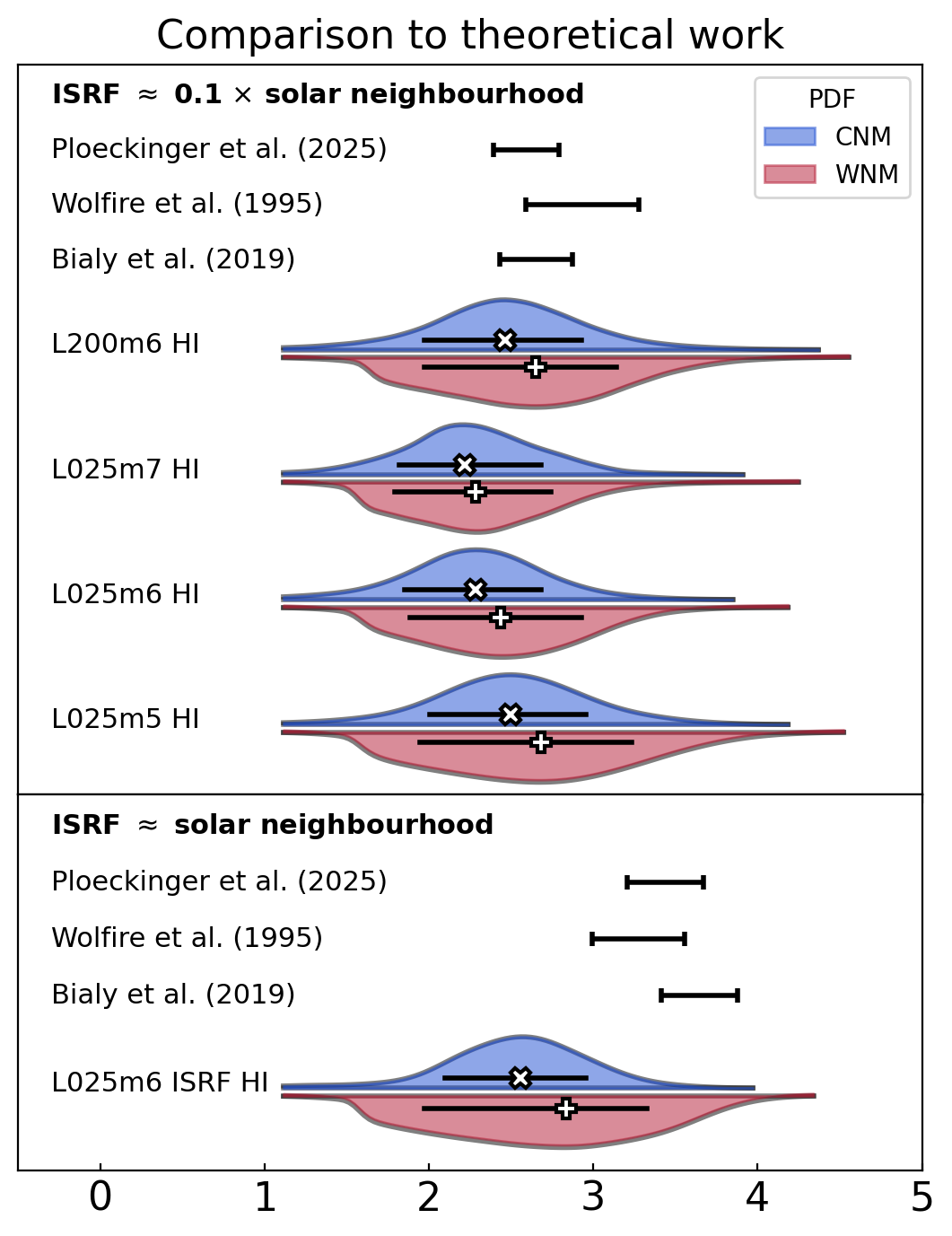}
    \caption{Thermal pressures of warm and cold neutral gas at solar metallicity for the ISRF strength used in \textsc{colibre} (top panel) and for a 10 times stronger ISRF, which is close to the solar-neighbourhood value (bottom panel).
    The pressure range for the multiphase ISM from theoretical work (black bars, \citealp{P25, Wolfire1995, Bialy2019})
    and the phase-separated PDFs weighted by \ion{H}{I} mass for galaxy samples in the L025m6 ISRF, L200m6, L025m7, L025m6 and L025m5 simulations from \textsc{colibre}, where maxima are marked by plus signs (WNM) and crosses (CNM). } 
    \label{fig:thermalpressures_literature}
\end{figure}

\subsubsection{Thermal pressures from theoretical work}\label{sec:disc:theory}

Theoretical work on the thermal pressures of the multiphase ISM typically relies on the comparison of radiative heating and cooling rates of gas in chemical steady-state and ionization equilibrium. 
For each gas density, the temperature at which these rates are in balance is defined as the equilibrium temperature, $T_{\mathrm{eq}}$ (see Appendix~\ref{sec:thermalequilibrium}). $T_{\mathrm{eq}}$, and subsequently $P_{\mathrm{eq}}$, generally do not include heating and cooling from hydrodynamic sources, such as turbulence or shocks. 
As shown in Figs.~\ref{fig:MW_hist_med_4} and \ref{fig:all_Z_hist_med_contour}, (see also \citealp{P25}) the thermal equilibrium curves provide a reference point for the thermal pressure of the gas in the simulated galaxies. 
The non-equilibrium species fractions (variations in relative elemental abundances), live-dust model (variations in dust properties), and additional heating processes in the simulations yield a variety of conditions resulting in significant scatter around the equilibrium curves.  

The two panels in Fig.~\ref{fig:thermalpressures_literature} compare theoretical thermal pressure ranges for the WNM and CNM to the thermal pressures of the \textsc{colibre} galaxies, both for ISRF intensities 0.1 times the solar-neighbourhood values (top panel) and for ISRF intensities consistent with the solar-neighbourhood values (bottom panel). These studies describe the general properties of neutral gas and we therefore compare them to the \ion{H}{I} mass-weighted PDFs of the \textsc{colibre} ISM phases. 

\citet{Wolfire1995} calculated $P_{\mathrm{eq}}(n)$ for the balance between various heating and cooling processes\footnote{ \citet{Wolfire1995} include PE heating from dust grains and polycyclic aromatic hydro-carbons (PAHs), CR heating, heating by soft X-rays and recombination cooling with the gas phase and on dust grains, metal-line cooling, and Ly$\alpha$ cooling.} and set a standard absorbing column density of $N_\mathrm{w} = 10^{19}\,\mathrm{cm}^{-2}$ (their notation), which is comparable to the shielding column density in \textsc{colibre} for typical temperatures and densities of the WNM. 
They further demonstrated that the pressure ranges, where WNM and CNM can coexist, shift to higher pressures for smaller shielding column densities, lower metallicities, higher dust abundances and stronger radiation fields (FUV and X-ray). 
The ISRF in \textsc{colibre} increases with gas density (see Sect.~\ref{sec:hybrid-chimes} for details), with a normalization of 0.1 times that of the solar neighbourhood value (``weakISRF'' model from \citealp{P25}; see Sect.~\ref{sec:disc:weakISRF} for a discussion). The lowest ISRF value in \citet{Wolfire1995} corresponds to a ISRF a factor of 0.3 lower than the integrated \citeauthor{Draine1978} field, resulting in $\log P_{\mathrm{min}}/k_{\rm B}\,[\mathrm{K\,cm}^{-3}] \approx 2.5$ and $\log P_{\mathrm{max}}/k_{\rm B}\,[\mathrm{K\,cm}^{-3}] \approx 3.2$. These are comparable to the thermal pressures of the WNM and CNM in solar metallicity \textsc{colibre} galaxies (see e.g.~Fig.~\ref{fig:MW_1D_pressure_hist} and the top panel of Fig.~\ref{fig:thermalpressures_literature}). 

\citet{Wolfire2003} repeated the analysis from \citet{Wolfire1995} with an updated chemical network and show that their $P_{\mathrm{min}}$ and $P_{\mathrm{max}}$ values increase with decreasing Galactocentric radii, due to the stronger radiation fields and increasing gas metallicities and dust abundances towards the Galactic centre. 
The example galaxies in \textsc{colibre} (Sect.~\ref{sec:MW_analogue}) exhibit the same radial dependence of thermal pressure and metallicity (Appendix~\ref{sec:appendix:radial}). 

The thermal equilibrium calculations of \citet{Bialy2019} include the same heating and cooling processes as in the model from \citet{Wolfire1995, Wolfire2003}, but they add cooling and heating from molecular hydrogen. Their model does not include attenuation of the radiation field and is thus only valid up to the edges of the CNM.
In their model H$_2$ is not important for the WNM-CNM phase transition at solar metallicities, in contrast with the \textsc{colibre} cooling model (see Appendix~\ref{sec:appendix:metdependenceexplanation}). 
For the same ISRF strength used in \textsc{colibre} ($I_{\mathrm{UV}} = 0.1$ in their notation), the WNM and CNM can be in pressure equilibrium between $\log P/k_{\mathrm{B}}\,[\mathrm{K\,cm^{-1}}] \approx 2.5$ and $\approx 2.9$ (top panel of Fig.~\ref{fig:thermalpressures_literature}). In their model, the metallicity-dependence of the pressure ranges of multiphase ISM is non-monotonic for metallicities around the solar value ($Z = 0.1 - 3\,\mathrm{Z}_{\odot}$). For lower metallicities, their pressures for the WNM-CNM coexistence increase with decreasing metallicity and the multiphased structure disappears for $Z\approx 10^{-5}\,\mathrm{Z}_{\odot}$. 

We focus our analysis on redshift $z=0$ and on well-resolved galaxies and gas densities, and are therefore limited to galaxies with average ISM metallicities of $Z_{\mathrm{ISM}}\ge0.1\,\mathrm{Z}_{\odot}$. In Fig.~\ref{fig:all_Z_hist_med_contour}, the thermal pressures of \textsc{colibre} galaxies exhibit a complex metallicity dependence, but we find higher median pressures in galaxies with $Z_{\mathrm{ISM}} = 0.1\,\mathrm{Z}_{\odot}$ than in those with $Z_{\mathrm{ISM}} = \,\mathrm{Z}_{\odot}$, in contrast to the findings of \citet{Bialy2019}. With PE heating as a critical heating source, this is likely related to the dust abundances, which are assumed to follow a broken power-law relation with metallicity in \citet{Bialy2019} while \textsc{colibre} uses the abundances from the live-dust model. Furthermore, in \textsc{colibre}, galaxies with different metallicities represent different types of galaxies, which is not captured in calculations in which metallicity is an independent parameter.  

To test whether a stronger radiation field increases the thermal pressures of WNM and CNM in \textsc{colibre} galaxies, a L025m6 simulation was run with an ISRF ten times stronger than that of the flagship simulation.  
For this model, the ISRF strength is close to that from \citet{Draine1978} at densities typical of the WNM - CNM transition. Our analysis shows that the thermal pressures in galaxies at solar metallicity are only slightly higher (see the WNM and CNM distributions in the bottom panel of Fig.~\ref{fig:thermalpressures_literature}, labelled ``L025m6 ISRF \ion{H}{I}'') than those in the standard \textsc{colibre} simulations (top panel in Fig.~\ref{fig:thermalpressures_literature}, labelled ``L025m6 \ion{H}{I}'').

\begin{figure} 
    \centering
    \includegraphics[width=0.98\linewidth]{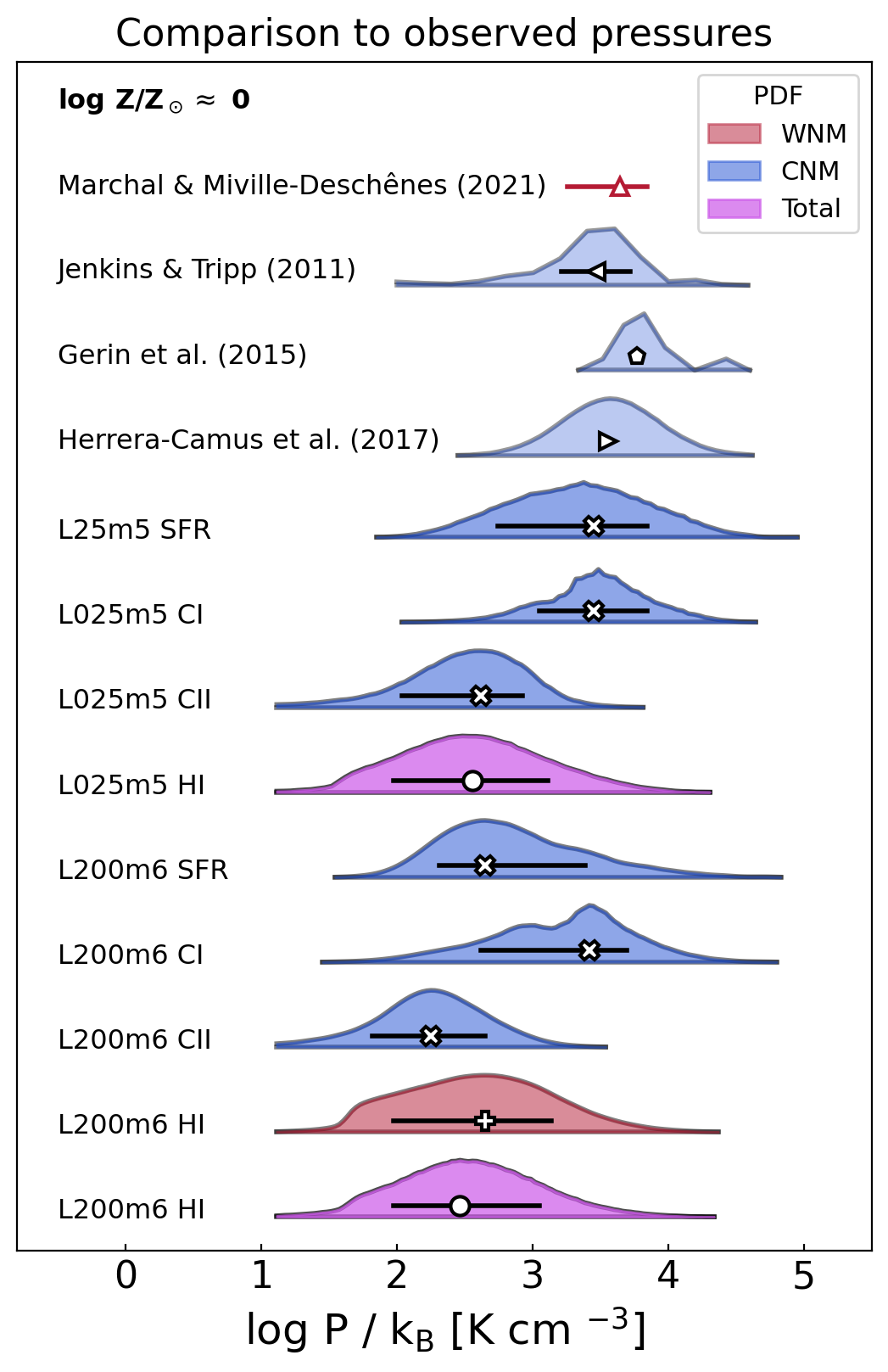}
    \caption{Thermal pressures of warm and cold neutral gas at a metallicity of $\log Z/\mathrm{Z}_{\odot} \approx 0$.
    The observed thermal pressures of warm (\citealp{MMD21}; red triangle) and cold (\citealp{JT11, Gerin2015, Herrera2017}; light blue distribution) neutral gas compared to the thermal pressure distributions for galaxy samples in the L025m5 and L200m6 simulations weighted by the SFR and the \ion{C}{I}, \ion{C}{II}, and \ion{H}{I} masses of the gas particles, where maxima are marked by plus signs (WNM) and crosses (CNM). }
    \label{fig:thermalpressures_obs_Z0.0}
\end{figure}

\begin{figure*}
    \centering
    \includegraphics[width=0.98\linewidth]{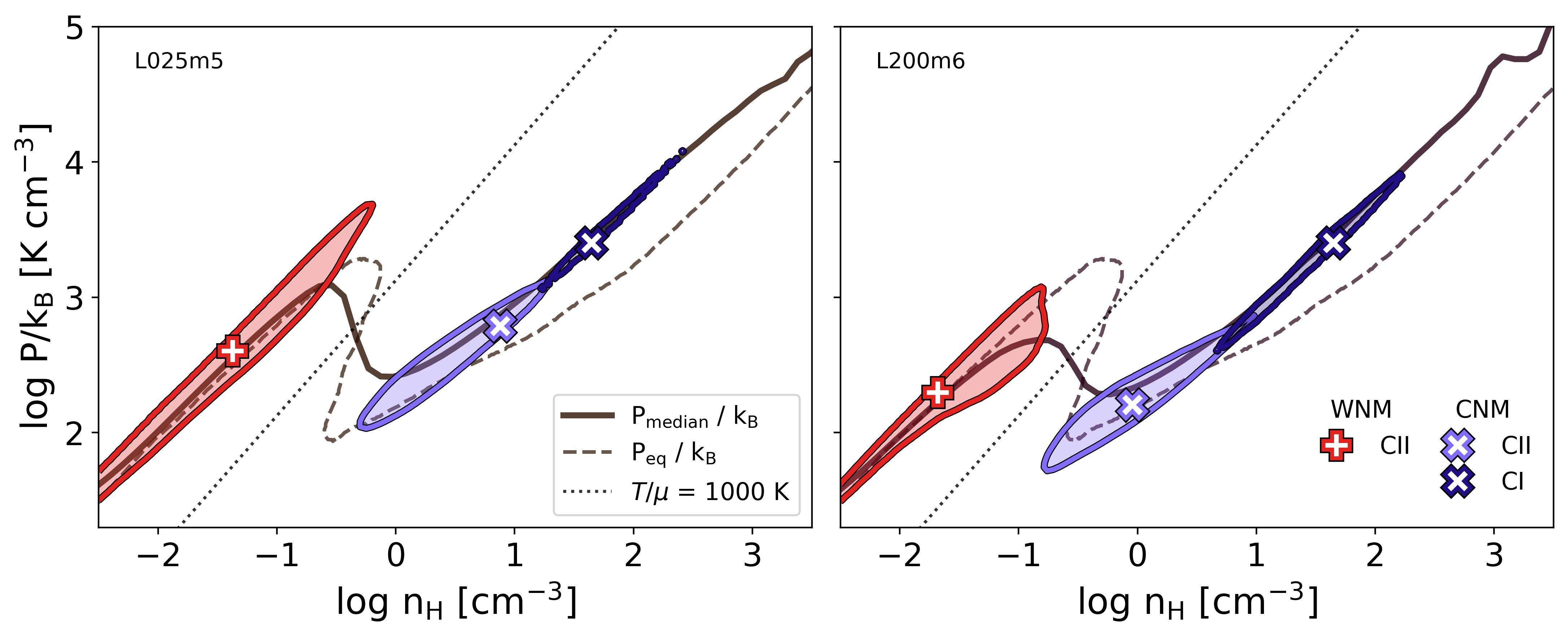}
    \caption{The thermal pressure $P/k_\mathrm{B}$ as a function of hydrogen number density $n_\mathrm{H}$ for galaxy samples with average ISM metallicity of $\log Z_\mathrm{ISM}/Z_\odot = 0\,\pm\, 0.1$ in the L025m5 (left panel) and L200m6 (right panel) simulation. The contours show the probability density function (containing $\approx 50\%$ of the mass of the respective phase) weighted by the the \ion{C}{II} mass (red/light blue) and the \ion{C}{I} mass (dark blue). The median (solid lines) and thermal equilibrium (dashed lines) pressures, as well as the temperature threshold for the CNM and WNM (dotted lines) are shown for reference. The maxima of the phase-separated PDFs are marked by plus signs (WNM) and crosses (CNM).
    } 
    \label{fig:CI_CII_HI_contours}
\end{figure*}

\subsubsection{Observed thermal pressures}\label{sec:disc:obs} 

In Fig.~\ref{fig:thermalpressures_obs_Z0.0}, we compare observed thermal pressures to the pressure distributions of \textsc{colibre} galaxies with a mean ISM metallicity close to the solar value. Observations generally derive the gas pressures from a combination of \ion{H}{I}, \ion{C}{I} and \ion{C}{II} data. For consistency, we compare these results to the \textsc{colibre} \ion{H}{I}, \ion{C}{I} and \ion{C}{II} mass-weighted PDFs. 
\ion{C}{II} emission is a dominant coolant in neutral gas and photodissociation regions, balancing the heating from young stars. Because of this the \ion{C}{II} emission is expected to trace the SFR (see e.g. \citealp{Sutter2019}, and references therein), which is why we include SFR weighted pressure distributions.
Note that we analyse large samples of simulated galaxies rather than specific objects, and we do not replicate observational techniques. This comparison is therefore meant as reference rather than a validation of the \textsc{colibre} ISM model.  

We compare to the Gaussian decomposition of \ion{H}{I} spectra performed by \citet{MMD21}. They measured the WNM column densities and combined these with dust extinction maps to extract the physical scales. They conclude that the derived average thermal pressure ($\log P/k_{\mathrm{B}} \,[\mathrm{K\,cm}^{-3}]= 3.64$) of their sampled region, which is located close to the edge of the Local Bubble at intermediate latitude, agrees with typical values for the solar neighbourhood. 
The thermal pressure distribution of the WNM in \textsc{colibre} galaxies (red distribution in Fig.~\ref{fig:thermalpressures_obs_Z0.0}) overlaps with the estimate by \citet{MMD21}, but the bulk of the simulated \ion{H}{I} mass lies at lower pressures.  

\citet{JT11} analysed \ion{C}{I} absorption lines in the spectra of 89 nearby stars with a method described in \citet{JenkinsShaya1979}. 
By construction, each sight-line includes a bright, early-type star which means that \citet{JT11} preferentially probe environments with stronger than average stellar radiation fields. 
They address this bias by analysing a subsample of sightlines with low radiation field intensities ($\log I/I_0 <0.5$, using the solar-neighbourhood reference value, $I_0$, from \citealp{Mathis1983}). They find that the mean thermal pressure of this subsample is $\log P/k_{\mathrm{B}}\, [\mathrm{K\,cm}^{-3}] = 3.47$, only slightly below the mean value of their full sample ($\log P/k_{\mathrm{B}}\, [\mathrm{K\,cm}^{-3}] = 3.58$). 
The \ion{C}{I} fraction increases with pressure and density \citep{JT11}. A small mass fraction of \ion{C}{I} in high-pressure gas along the line of sight is very prominent, making \ion{C}{I} biased towards higher pressures. 
The \ion{C}{I} mass-weighted pressure distributions of the galaxy samples in the L200m6 and L025m5 simulations (Fig.~\ref{fig:thermalpressures_obs_Z0.0}) shift to higher pressures compared to the \ion{H}{I} mass-weighted distributions and coincide with the results from \citet{JT11}. The \ion{C}{I} mass is concentrated in the CNM and occupies high densities in \textsc{colibre} galaxies (dark blue contours in Fig.~\ref{fig:CI_CII_HI_contours}). 

\citet{Gerin2015} used measurements of the fine-structure line of singly-ionized carbon\footnote{Singly-ionised carbon is also present in neutral hydrogen gas, because neutral carbon has a lower ionization potential than hydrogen.}, [\ion{C}{II}] at $158\,\mu \rm m$, in combination with \ion{C}{I} and \ion{H}{I} data to derive the thermal pressures of diffuse gas along sightlines towards bright dust continuum regions in the Galactic plane. 
For a mean kinetic temperature of $100\,\mathrm{K}$ (from the \ion{H}{I} Galactic plane survey, \citealp{McClure2009, Kalberla2010}), they found that the measured \ion{C}{II} and \ion{C}{I} excitation in the diffuse gas is consistent with a median pressure of $\log P/k_{\mathrm{B}} \, [\mathrm{K\,cm}^{-3}] \approx 3.8$.  

With a similar approach, \citet{Herrera2017} analysed [\ion{C}{II}], \ion{H}{I} and CO data in 534~regions (size $\approx 1\,\mathrm{kpc}$) of predominately atomic gas in 31~galaxies from the \textit{Herschel} KINGFISH sample \citep{Kennicutt2011PASP}. 
They find log-normal distributions with median values ranging from $\log P/k_{\mathrm{B}} \,[\mathrm{K\,cm}^{-3}]= 3.4$ to $3.8$. 
\cite{Herrera2017} additionally impose a threshold on the mean intensities ($\langle U \rangle \leq 3$ in their notation) and find a median thermal pressure of $\log P/k_{\mathrm{B}} \,[\mathrm{K\,cm}^{-3}]= 3.6$, which coincides with the estimates for the Galactic plane from \cite{JT11} and \cite{Gerin2015}. 

The Galactic [\ion{C}{II}] emission emerges from dense photodissociation regions ($30~\%$), CO-dark gas ($25~\%$), cold \ion{H}{I} ($25~\%$) and ionized gas ($20~\%$) \citep{Pineda2014}. 
With strong FUV radiation from star formation associated with some of these regions, the \ion{C}{II} emission is correlated with the star formation rate \citep[see e.g.,][]{Herrera2015}. Collisional excitation of carbon ions (by electrons, H, and H$_2$) is more likely in regions with higher pressures and densities \citep{Herrera2017}. 
Thermal pressures derived from \ion{C}{II} emission may therefore be biased towards higher pressures. The studies mentioned here address these biases, but they acknowledge that their estimates may still be skewed towards higher pressures. 

The SFR weighted and \ion{C}{I} mass-weighted pressure distributions of the solar-metallicity galaxy sample in the L025m5 simulation match the estimates from \cite{Gerin2015} and \cite{Herrera2017}. The SFR-weighted distributions are more relevant for estimates based on \ion{C}{II} emission than the \ion{C}{II} mass-weighted ones, since \ion{C}{II} emission is expected to trace the SFR.
The \ion{C}{II} mass-weighted pressure distribution peaks at a thermal pressure of $\log P/k_{\mathrm{B}} \,[\mathrm{K\,cm}^{-3}]\approx 3$ (L025m5), which is $\approx 1~\mathrm{dex}$ lower than the SFR weighted and \ion{C}{I} mass-weighted distributions (Fig.~\ref{fig:thermalpressures_obs_Z0.0}). 
Galaxies simulated at m6 resolution yield lower thermal pressures than at m5 resolution. The \ion{H}{I}, \ion{C}{I}, and \ion{C}{II} mass-weighted distributions of galaxies in the L200m6 simulation deviate only slightly from those of galaxies in the L025m5 simulation.
The SRF weighted pressure distribution of galaxies in the L200m6 simulation peaks at $\approx 1~\rm dex$ lower values than that of galaxies in the L025m5 simulation. This may be attributed to the difference in resolution and possibly the difference in sample size. 

The thermal pressure distributions are pushed to higher pressures when weighted by the SFR or the \ion{C}{I} mass of the gas particles, indicating the bias of observational estimates towards high pressure regions. 
We need mock \ion{C}{I} absorption and \ion{C}{II} emission line data to confirm that observational estimates preferentially trace high pressure environments.

The contribution of the neutral phases to the \ion{C}{II} emission is an active debate. For typical conditions of the neutral phases in the MW\footnote{With volume densities and temperatures of the CNM ($n_\mathrm{H} = 50~\mathrm{cm}^{-3}, T = 100~\mathrm{K}$) and WNM ($n_\mathrm{H} = 0.5~\mathrm{cm}^{-3}, T = 8000\, \mathrm{K}$).}, the CNM contributes $\approx 20$ times more to the \ion{C}{II} emission than the WNM \citep{Herrera2017}. 
In the solar-metallicity galaxy sample of the L200m6 simulation (Fig.~\ref{fig:CI_CII_HI_contours}), the \ion{C}{II} mass is more evenly distributed between the CNM (with $\approx 45~\%$ of the total \ion{C}{II} mass) and WNM ($\approx 55~\%$). In contrast to the \ion{C}{I} mass that is predominantly found in the CNM ($\approx 94~\%$). Similar results can be seen for galaxies in the L025m5 simulation.

\section{Summary}\label{sec:summary}

    We present the multiphase structure of the atomic ISM in galaxies at redshift $z=0$ from the \textsc{colibre} cosmological simulations \citep{Schaye2025colibre, Chaikin2025calibration}. The multiphase ISM in \textsc{colibre} is an emergent property arising from allowing gas to cool down to $\approx10~\mathrm{K}$, using the \textsc{hybrid-chimes} framework \citep{P25} based on the \textsc{chimes} chemical network \citep{Richings2014optthin, Richings2014shielded} for the radiative cooling and heating rates, and a live-dust model \citep{Trayford2025} that follows the evolution of dust grains. The non-resolved \ion{H}{II} regions from the early feedback model \citep{BenitezLlambay2026} do not interfere greatly with the effective properties of the ISM reported in this work.
    We primarily use the \textsc{colibre} simulations with both baryonic and dark matter particle masses of the order of $10^{5}\,\mathrm{M}_{\odot}$ (m5) and $10^{6}\,\mathrm{M}_{\odot}$ (m6) and the largest volumes at each resolution (Table~\ref{tab:colibre_volumes}). 
    
    The \textsc{colibre} simulations produce the multiphase structure of the neutral atomic ISM, with a clear separation into WNM and CNM that generally follows the typical S-shape seen in the thermal equilibrium curve of pressure versus density. 
    The thermal pressure distributions of the neutral phases in \textsc{colibre} galaxies shows a dependence on the average metallicity of galaxies. We discuss how the strength of the interstellar radiation field (ISRF) in the chemical model, the dust abundance from the live-dust model, and the resolution influence the thermal pressure of the ISM. We offer a comparison against predictions from theoretical models and observations in solar metallicity environments.
    Our results can be summarized as follows: 

    \begin{enumerate}
        \item Four example galaxies with roughly solar metallicity taken from the L025m5 simulation show a well-defined multiphase structure (Fig.~\ref{fig:MW_hist_med_4}). The WNM and CNM coexist within an approximate pressure range between $\log P_{\mathrm{min}}/k_{\mathrm{B}}\,[\mathrm{K\,cm}^{-3}]\approx 2$ and $\log P_{\mathrm{max}}/k_{\mathrm{B}}\,[\mathrm{K\,cm}^{-3}]\approx 3$, taken from \ion{H}{I} mass-weighted distributions (Fig.~\ref{fig:MW_1D_pressure_hist}).
        
        \item The multiphase pressure range is similar for \ion{H}{I} gas in large galaxy samples in narrow metallicity bins with $\log Z_{\mathrm{ISM}}/Z_\odot \ge -0.5$ (Fig.~\ref{fig:all_Z_hist_med_contour}), taken from the L200m6 \textsc{colibre} simulation, and shifts to higher pressures for galaxies with metallicities of $\log Z_{\mathrm{ISM}}/Z_\odot = -1$ (Fig.~\ref{fig:Pk_1D_hists}). 
        
        \item We highlight that the ISRF in \textsc{hybrid-chimes} varies with temperature and density (Fig.~\ref{fig:G0}). The thermal pressure distribution of \textsc{colibre} galaxies with solar metallicity is shifted to slightly higher pressures when the ISRF is increased by a factor of 10 (compare ``L025m6 \ion{H}{I}'' and ``L025m6 ISRF \ion{H}{I}'' in Fig.~\ref{fig:thermalpressures_literature}). This pressure increase is comparable to that from increasing the mass resolution of the simulation by a factor of eight (compare ``L025m6 \ion{H}{I}'' and ``L025m5 \ion{H}{I}'' in Fig.~\ref{fig:thermalpressures_literature}).

        \item On average, the $\mathcal{DTG}$ in the simulated galaxies with solar metallicity (Fig.~\ref{fig:DTZ_solar_Z_bin}) is below the equilibrium value in diffuse gas and increases to match it in dense gas, as a result of the live-dust model. The lower $\mathcal{DTG}$ shifts the balance between heating and cooling rates towards lower equilibrium temperatures, offering an explanation for the difference between the thermal pressures of \ion{H}{I} gas in \textsc{colibre} galaxies and the thermal equilibrium pressures. 

        \item The thermal pressures of the neutral phases in \textsc{colibre} are not converged with the numerical resolution. The median thermal pressures of galaxy samples at different metallicities in the L025m7, L025m6, and L025m5 simulations (Fig.~\ref{fig:Compare_res_L25_Z_med}) increase with increasing resolution at densities of the WNM - CNM transition.
        
        \item The thermal pressure ranges of the multiphase ISM from theoretical works \citep{Wolfire1995, Bialy2019} are comparable to the thermal pressure distributions of the warm and cold phases in \textsc{colibre} (Fig.~\ref{fig:thermalpressures_literature}), for solar metallicity and similar ISRF strengths ($\sim 0.1$ times the solar neighbourhood value).  

        \item The thermal pressure distributions weighted by the SFR or \ion{C}{I} masses of particles in the CNM of galaxies with solar metallicity in the L025m5 simulation (Fig.~\ref{fig:thermalpressures_obs_Z0.0}) match well with observational estimates \citep{Gerin2015, Herrera2017, JT11}. 
        The \ion{C}{II} mass-weighted thermal pressure distributions of the simulated galaxies have $\approx 1~\mathrm{dex}$ lower pressures than observational studies using \ion{C}{II} emission data (in combination with \ion{H}{I} and \ion{C}{I} or CO data; \citealp{Gerin2015, Herrera2017}), where the difference is smaller for the higher resolution simulation. 
        The SFR weighted distributions are more relevant than the \ion{C}{II} mass-weighted ones, since \ion{C}{II} emission is correlated with the SFR. 
        
        \item We expect \ion{C}{II} and especially \ion{C}{I} data to preferentially trace higher pressure and density environments. The \ion{C}{I} mass in \textsc{colibre} is concentrated in the CNM and occupies high pressures and densities, matching the expectation from observations, while the \ion{C}{II} mass is more evenly distributed between the warm and cold neutral phases and sits at lower pressures and densities (Fig.~\ref{fig:CI_CII_HI_contours}). 
    \end{enumerate}
   
    We conclude that a combination of the ISRF strength, variable $\mathcal{DTG}$ from the live-dust model, and the resolution are responsible for the difference between the thermal pressures of the WNM and CNM in \textsc{colibre} galaxies at solar metallicity to those from theoretical models. 
    As we show in Fig.~\ref{fig:Compare_res_L25_Z_med}, simulated galaxies with higher resolution have higher thermal pressures. We need higher resolution simulations to study the thermal pressures of the neutral phases, particularly since they better resolve cold gas at lower metallicities (Fig.~\ref{fig:all_Z_hist_med_contour}). 
    
    Our approach to a more realistic comparison with observations, where the thermal pressure distributions are weighted by the chemical species that are used as tracers in the observational studies, shows the \ion{C}{I} mass-weighted and SFR weighted distributions agree with observational estimates, which implies that \ion{C}{I} and \ion{C}{II} data is biased to higher pressure environments. Note that SFR weighting is most appropriate for a comparison to studies using \ion{C}{II} emission line data. 

    The large differences between the thermal pressures derived with different weighting schemes (Fig~\ref{fig:thermalpressures_obs_Z0.0}) highlights the importance of virtual observations for a more direct comparison.
    Recreating observational techniques involves performing radiative transfer (using codes such as \textsc{radmc-3d} \citealp{Dullemond2012}, \textsc{rascas} \citealp{MichelDansac2020} or \textsc{skirt} \citealp{CampsBaes2015, CampsBaes2020}) and photoionization (with codes like \textsc{cloudy} \citealp{Ferland1998, Chatzikos2023}) in post-processing of individual galaxies to produce mock absorption and emission line data (e.g. \citealp{Padilla2021, Padilla2023, Gazagnes2024, Chan2026}), to investigate whether the differences between observational and simulated findings stem from methodology or physical processes. 

    Throughout this work, we highlight the limitations of thermal equilibrium models in estimating the physical properties of the stable phases in the ISM (see also the discussion in \citealp{P25}), and conclude that, while a valid reference point to investigate the underlying processes, care should be taken in the interpretation of equilibrium models.

\begin{acknowledgements}
    MH is grateful to Lilly Kormann and Michelle Anderson for helpful discussions.
    
    SP acknowledges support by the Austrian Science Fund (FWF) through grant-DOI: 10.55776/V982. 
    EC acknowledges support from STFC consolidated grant ST/X001075/1.
    ABL acknowledges support by the Italian Ministry for Universities (MUR) program “Dipartimenti di Eccellenza 2023-2027” within the Centro Bicocca di Cosmologia Quantitativa (BiCoQ), and support by UNIMIB’s Fondo Di Ateneo Quota Competitiva (project 2024-ATEQC-0050).
    FH acknowledges funding from the Netherlands Organization for Scientific Research (NWO) through research programme Athena 184.034.002.

    We acknowledge the use of python packages, including the publicly available \textsc{swiftsimio} \citep{Borrow2020, Borrow2021} and \textsc{swiftgalaxy} \citep{Oman2025, Oman2025swiftgalaxy} modules.

    This work used the DiRAC@Durham facility managed by the Institute for Computational Cosmology on behalf of the STFC DiRAC HPC Facility (\url{www.dirac.ac.uk}). The equipment was funded by BEIS capital funding via STFC capital grants ST/K00042X/1, ST/P002293/1 and ST/R002371/1, Durham University and STFC operations grant ST/R000832/1. DiRAC is part of the National e-Infrastructure.
\end{acknowledgements}

\section*{Data Availability}
    The figures resulting from our analysis of galaxies with solar metallicity within the L025m5 simulation that meet the selection criteria in Sect.~\ref{sec:MW_analogue} are available here\footnote{To be added upon publication.}.

    The \textsc{swift} code \citep{Schaller2024swift} is publicly available at \href{www.swiftsim.com}{\texttt{swiftsim.com}}. The \textsc{colibre} galaxy formation module integrated in \textsc{swift} will be available after the public release of the simulations, for more information visit the \textsc{colibre} webpage at \href{www.colibre-simulations.org}{\texttt{colibre-simulations.org}}. 
    
    The \textsc{hybrid-chimes} project webpage is \href{https://www.sylviaploeckinger.com/hybridchimes}{\texttt{sylviaploeckinger.com/hybridchimes}} and the chemical network \textsc{chimes} \citep{Richings2014optthin,Richings2014shielded} is publicly available at \href{https://richings.bitbucket.io/chimes/download.html}{\texttt{richings.bitbucket.io/chimes}}.

%
\bibliographystyle{aa} 
\bibliography{aa} 

\begin{appendix}
\nolinenumbers





\section{Thermal equilibrium}\label{sec:thermalequilibrium}

    \begin{figure}
        \centering
        \includegraphics[width=\linewidth]{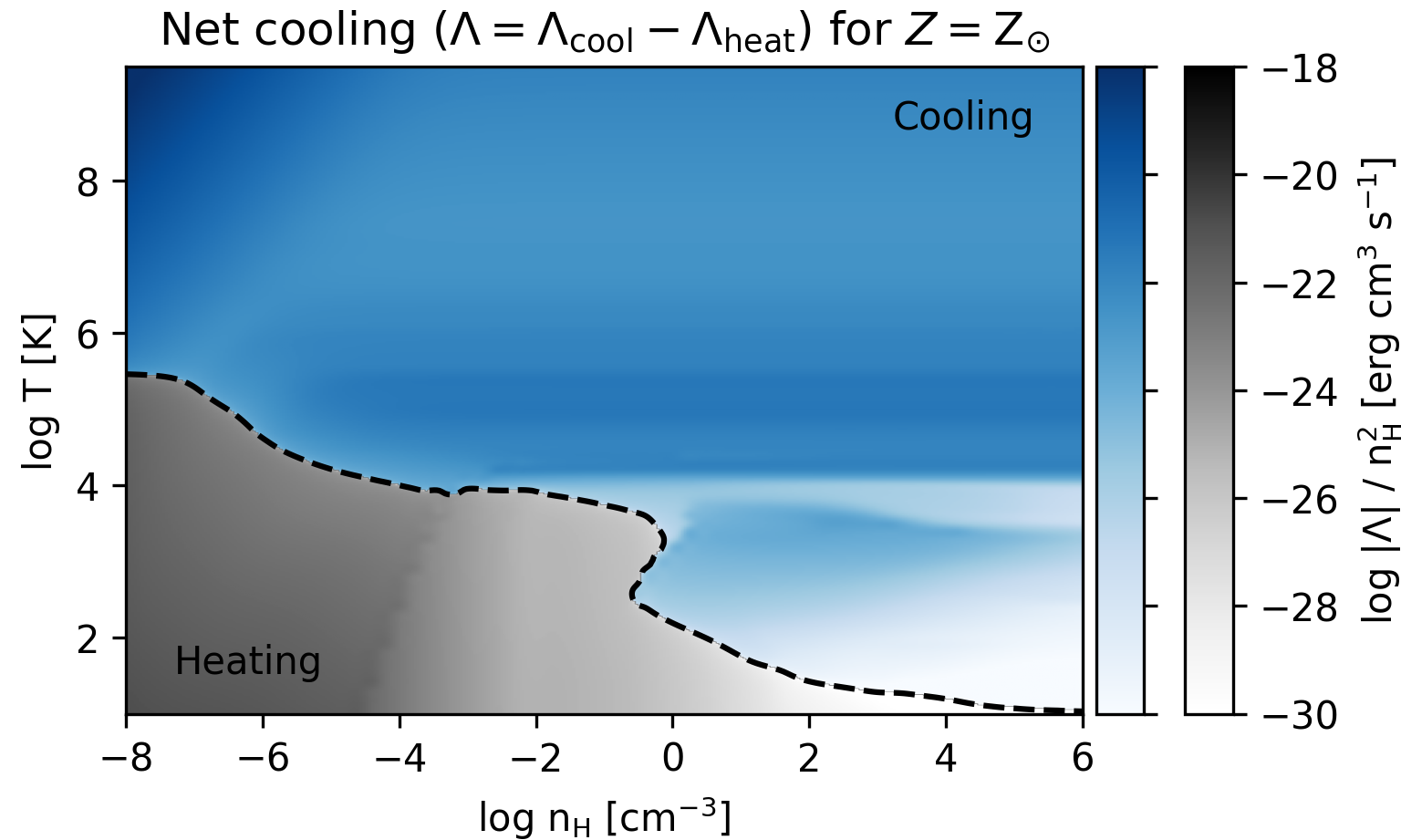}
        \caption{Net cooling rates $\log |\Lambda| / n_{\mathrm{H}}^2 \,[\mathrm{erg\,cm}^{3}\,\mathrm{s}^{-1}]= \log |\Lambda_{\mathrm{cool}} - \Lambda_{\mathrm{heat}}| / n_{\mathrm{H}}^2$ of solar metallicity gas, using the same assumptions about the ISRF, CR rate, shielding column density and UV background radiation (here for $z=0$), as in \textsc{colibre} but for chemical equilibrium and a constant dust-to-gas ratio ($\mathcal{DTG} = 6.6\times10^{-3}$). The thermal equilibrium contour (dashed line) indicates the heating and cooling balance. Cooling (heating) dominates at temperatures above (below) the dashed line. }
        \label{fig:Teqexample}
    \end{figure}

To facilitate the comparison with the multiphase structure of the ISM predicted by thermal equilibrium calculations and a more in-depth analysis of the involved processes, we construct equilibrium functions directly from the tabulated cooling and heating rates. The rates are calculated with the \textsc{hybrid-chimes} pipeline\footnote{The python routines to create cooling and heating rate tables are publicly available at \href{https://gitlab.phaidra.org/hybridchimes/hybridchimes}{\texttt{gitlab.phaidra.org/hybridchimes/hybridchimes}}.} under the same assumptions for the ISRF, the CR rate, the shielding column density, and the same chemical network as in \textsc{colibre}. Nonetheless, a few differences between our equilibrium functions and the radiative cooling model in \textsc{colibre} remain and we describe here the corrections that are applied to the tabulated rates to allow for a more direct comparison with the simulation output. 

First, in \textsc{colibre}, the cooling and heating rates from hydrogen and helium species are calculated from non-equilibrium species abundances. These abundances are evolved on the fly by \textsc{chimes}, integrated into \textsc{swift}. As shown by \citet{Ploeckinger2025_H2}, a reduced chemical network that does not include oxygen may overestimate the H$_2$ fractions, which increases H$_2$ cooling. For a better match with the \textsc{colibre} chemistry calculations, we tabulate the equilibrium species abundances for all included elements (H, He, C, N, O, Ne, Mg, Si, S, Ca, and Fe), but replace those from H and He species with their abundances from a reduced network (containing only H and He). The cooling and heating rates are then calculated based on these combined abundances. 

Second, the tabulated rates do not include any dust depletion by default. We correct the metal cooling rates by the depletion factors of \citet{Jenkins2009} for their strong depletion model, parametrized by $F_{\star} = 1$. In contrast, \textsc{colibre} utilizes the live-dust model of \cite{Trayford2025}, which follows the depletion of elements from the gas phase onto dust grains self-consistently. We will discuss the effects of the live-dust model on the dust-to-gas ratio in Sect.~\ref{sec:disc:dust}. 

Fig.~\ref{fig:Teqexample} shows an overview of the tabulated net cooling rates $\log |\Lambda| / n_{\mathrm{H}}^2 \,[\mathrm{erg\,cm}^{3}\,\mathrm{s}^{-1}]= \log |\Lambda_{\mathrm{cool}} - \Lambda_{\mathrm{heat}}| / n_{\mathrm{H}}^2$ for solar metallicity gas. The temperature - density space is split by the thermal equilibrium temperature ($T_{\mathrm{eq}} \equiv T_{\Lambda = 0}$; dashed line) into a cooling dominated (high temperature) and a heating dominated (low temperature) part. Note that the thermal equilibrium temperature is multi-valued close to $\log n_{\mathrm{H}}\,[\mathrm{cm}^{-3}] \approx -0.3$. This is caused by a strong peak in the H$_2$ cooling rate close to a temperature of $\log T\,[\mathrm{K}] = 3$ (see Appendix~\ref{sec:appendix:metdependenceexplanation}). The thermal equilibrium pressure, $P_{\mathrm{eq}}$, is defined as

\begin{equation}
    P_\mathrm{eq}/k_\mathrm{B} = \frac{1}{X_{\mathrm{H}}\mu_{\mathrm{eq}}} \, n_{\mathrm{H}} \,T_{\mathrm{eq}} \quad ,
    \label{eq:eq_pressure}
\end{equation}

\noindent
where $k_{\mathrm{B}}$ is the Boltzmann constant, $X_{\mathrm{H}}$ is the hydrogen mass fraction, $\mu_{\mathrm{eq}}$ is the mean particle mass in units of the proton mass at $T_{\mathrm{eq}}$, and $n_{\mathrm{H}}$ is the hydrogen number density.

For the gas particles in a \textsc{colibre} galaxy, the picture is more complex: at a given point in temperature-density space, the species abundances and the dust abundance, and therefore the heating and cooling rates, depend on the past history of each particle. Despite the differences from the full \textsc{colibre} cooling model, (i.e. the non-equilibrium species abundances of H and He and the live-dust abundances), we use these equilibrium temperatures (and equilibrium pressures, $P_{\mathrm{eq}}$) as a reference when discussing the thermal pressures of the ISM in \textsc{colibre} galaxies.

\section{Radial dependence of thermal pressures}\label{sec:appendix:radial}

    \begin{figure*}
        \centering
        \includegraphics[width=1.0\linewidth]{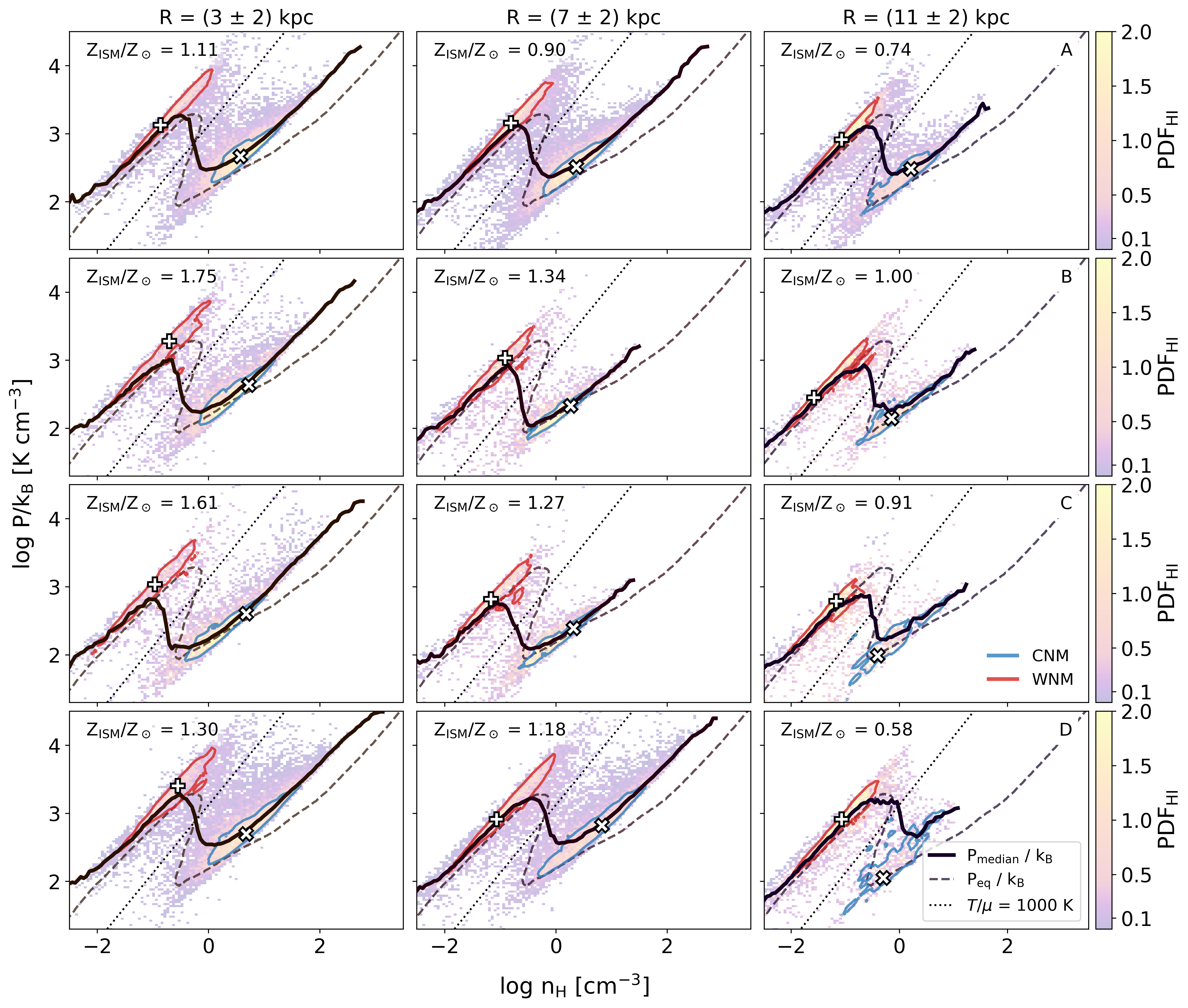}
        \caption{
        The thermal pressure $P/k_\mathrm{B}$ as a function of hydrogen number density $n_\mathrm{H}$ for \ion{H}{I} gas within three radial bins, centred at radii $R =$ 3 (column one), 7 (column two), and 11 (column three) $\pm 2$~kpc. Each row shows one of the example galaxies A through D from top to bottom, respectively. The components of each panel are the same as in Fig.~\ref{fig:MW_hist_med_4} but for the particles in each annuli. 
        }
        \label{fig:apertures_3_each_row_MW} 
    \end{figure*}
    
    For the four example galaxies (Sect.~\ref{sec:MW_analogue}), we select a cylindrical region extending $\pm5\, \mathrm{kpc}$ vertically above the plane of each galaxy and select gas particles within three different galactocentric radii, $R\,\pm 2~\rm kpc$. The values for $R$ represent the inner part of the galaxy ($R = 3\,\mathrm{kpc}$), an intermediate region ($R=7\,\mathrm{kpc}$), and the outskirts of the galaxy ($R = 11\,\mathrm{kpc}$). Note that the gas mass surface densities at these radii can vary greatly (see Fig.~\ref{fig:Radial_surf_dens_MW}). 
    
    The distributions of \ion{H}{I} gas in the WNM and CNM at different radii of all four example galaxies with approximately solar metallicity ISM are shown with red and blue contours in Fig.~\ref{fig:apertures_3_each_row_MW}; the PDF peaks are marked as in Fig.~\ref{fig:MW_hist_med_4}. The equilibrium pressure for solar metallicity (dashed line) is shown for orientation and not a prediction for gas at these metallicities.
    The typical pressures of both the CNM and WNM decrease with increasing galactocentric radius in all four galaxies. This is qualitatively consistent with the model from \citet{Wolfire2003}, in which the transition from warm to cold gas also shifts to lower pressures for larger galactocentric radii (see Sect.~\ref{sec:discussion_lit} for details). In their work, the radiation field, cosmic ray rates, gas phase metallicities, and dust abundances depend on galactocentric distance. 
    The radial dependence of these processes in the \textsc{colibre} galaxies are calculated self-consistently resulting from the scaling of the radiation field and cosmic ray ionization rates with the local gas density and temperature (see \citealp{P25} for details), as well as from stellar enrichment \citep{Correa2026}, the live-dust evolution model \citep{Trayford2025}, and to a small degree from the implementation of \ion{H}{II} regions \citep{BenitezLlambay2026}.

    The WNM and CNM pressure range changes only moderately between the inner two radial bins. The CNM in the intermediate region (middle column) is largely consistent with the cold gas distribution of the full disk (compare with Fig.~\ref{fig:MW_hist_med_4}), whereas the peak of the WNM PDF is located at higher pressures closer to the centre. 
    Note that the galaxy exhibits a negative metallicity gradient with galactocentric radius. The difference in gas metallicity from the central bin to the outermost radial bin is 0.18, 0.24, 0.25, and 0.35~$\mathrm{dex}$, for galaxies A, B, C, and D, respectively. Most of the \ion{H}{I} gas in the outskirts of the galaxies is in the WNM.

\section{Metallicity-selected galaxy samples}\label{sec:appendix:metallicity_dependence}

    \begin{figure*}
        \centering
        \includegraphics[width=0.95\linewidth]{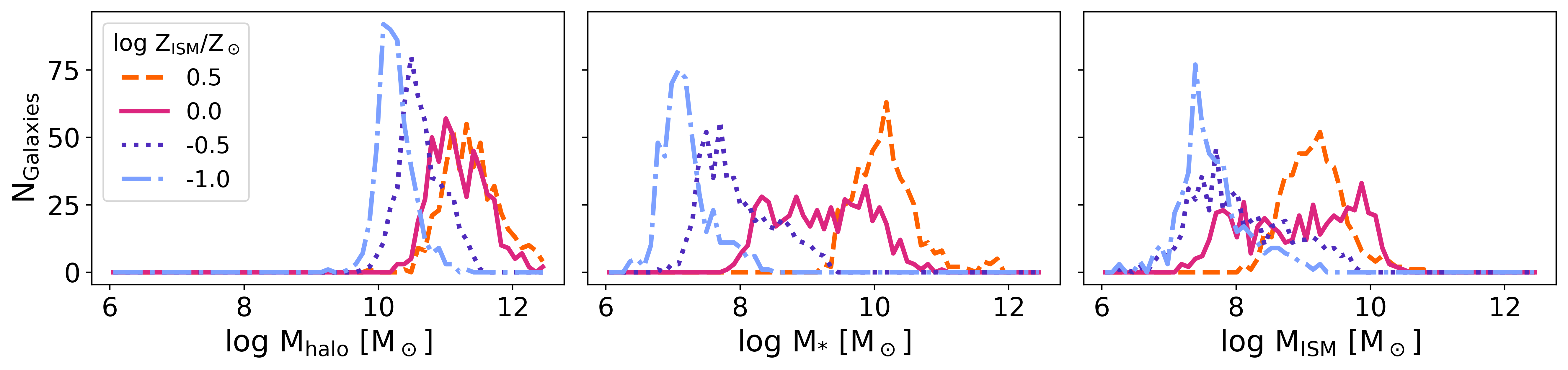}
        \caption{Distribution of halo masses $M_\mathrm{halo}$ (first panel), stellar masses $M_\star$ (second panel), and cool ($T<10^{4.5}\,\mathrm{K}$), dense ($n_{\mathrm{H}}>0.1\,\mathrm{cm}^{-3}$) gas masses $M_\mathrm{ISM}$ (third panel) for all 500 galaxies in each metallicity bin in the L200m6 simulation. The bin width in masses is $0.1~\mathrm{dex}$. 
        }
        \label{fig:mass_hists} 
    \end{figure*}

In Fig.~\ref{fig:mass_hists}, we show the halo masses (left), stellar masses (middle), and ISM gas masses (right) of the metallicity-selected sample of galaxies in Sect.~\ref{sec:metallicity_dependence}. While bins with increasing ISM metallicity represent galaxies with increasing average halo, stellar, and ISM masses, there is a significant scatter and overlap in these properties.

\section{What drives the metallicity dependence in \textsc{colibre}?}\label{sec:appendix:metdependenceexplanation}

\begin{figure} 
    \includegraphics[width=0.98\linewidth]{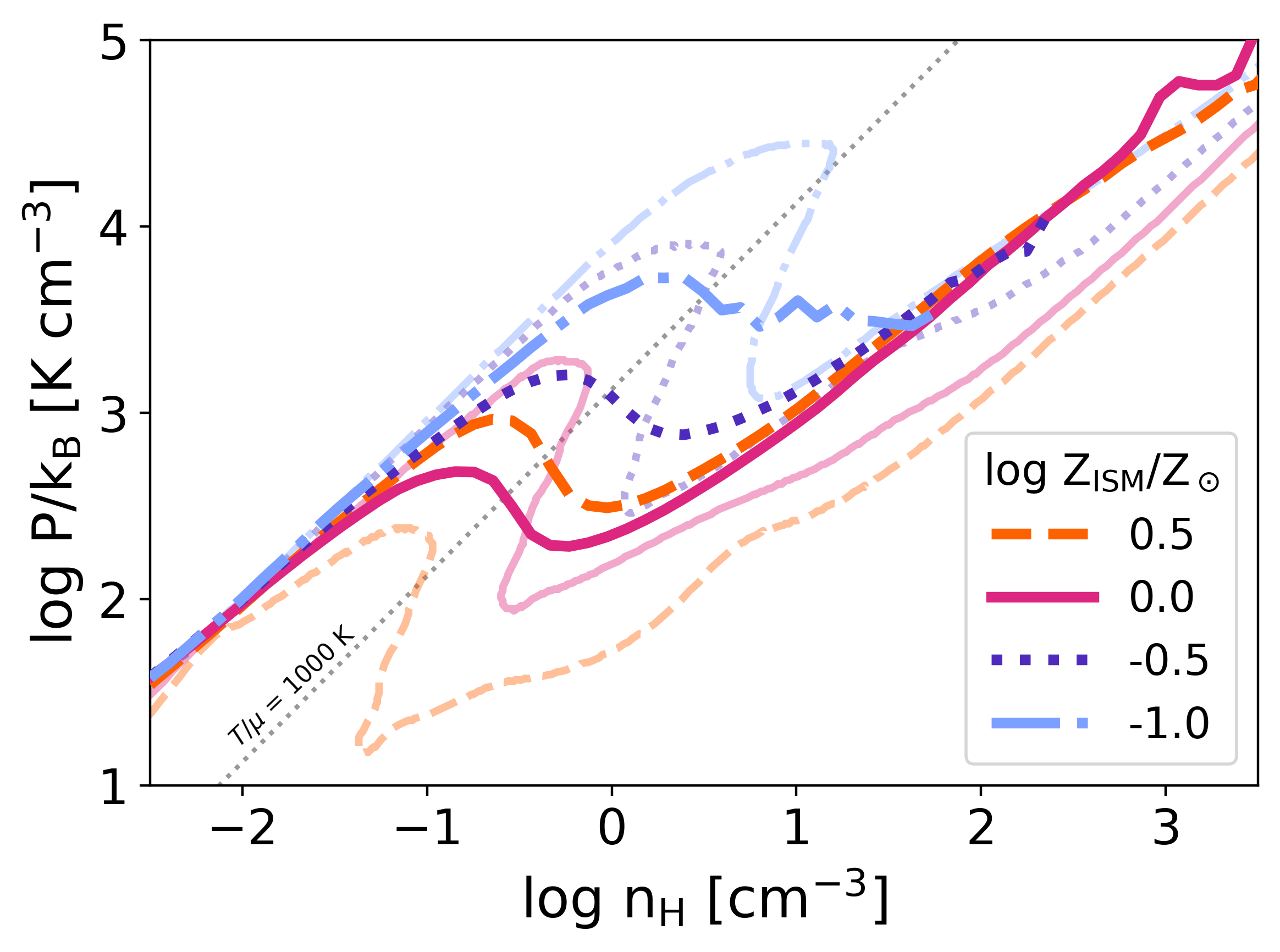}
    \caption{The median thermal pressure $P_\mathrm{median}/k_\mathrm{B}$ (opaque lines) and equilibrium thermal pressure $P_\mathrm{eq}/k_\mathrm{B}$ (transparent lines) as a function of hydrogen number density $n_\mathrm{H}$ for all metallicity bins.} 
    \label{fig:all_Z_Pk_T} 
\end{figure}

\begin{figure}
    \includegraphics[width=0.98\linewidth]{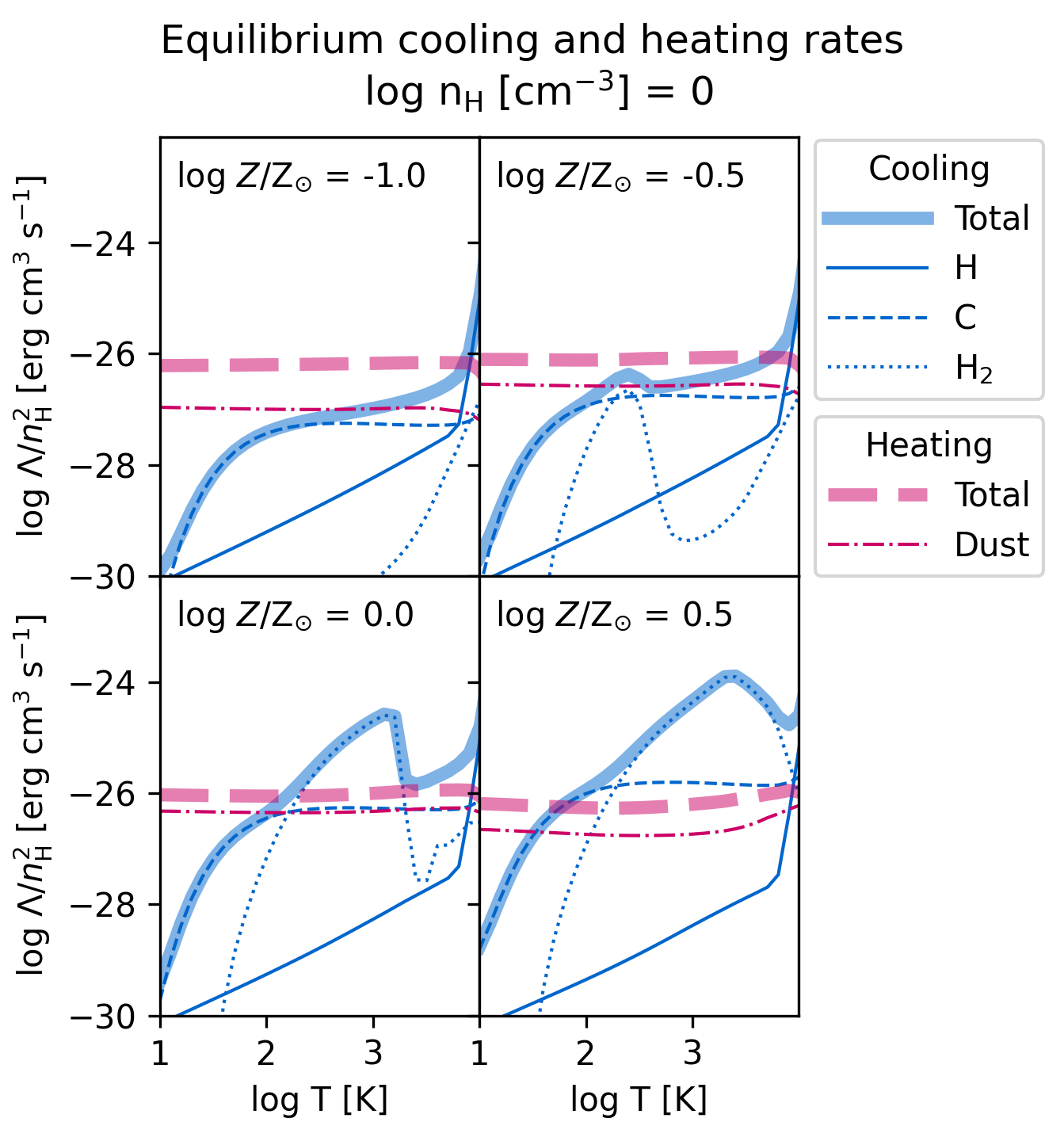}
    \caption{Selected equilibrium cooling and heating rates for gas with a density of $\log n_{\mathrm{H}}\,[\mathrm{cm}^{-3}] = 0$ and metallicities from $\log Z/\mathrm{Z}_{\odot} = -1$ (top left panel) to $0.5$ (bottom right panel). Temperatures at which the total cooling rate (thick solid line) balances the total heating rate (thick dashed line), are referred to as equilibrium temperatures, $T_{\mathrm{eq}}$, which decreases $T_{\mathrm{eq}}$ with increasing metallicity. The total cooling is dominated by cooling from carbon (\ion{C}{II}: thin dashed line) and hydrogen species (H$_2$ with a peak at $\log T\,[\mathrm{K}]\approx 3$: dotted line; and \ion{H}{I} at $\log T\,[\mathrm{K}]\approx 4$ through Ly$\alpha$ emission; thin solid line). Dust heating (dash-dotted line) is an important heating process. } 
    \label{fig:rates_logn0}
\end{figure}

Fig.~\ref{fig:all_Z_Pk_T} summarizes the metallicity dependence of the multiphase ISM pressures in \textsc{colibre} galaxies (Sect.~\ref{sec:metallicity_dependence}). The opaque lines represent the median pressures of \ion{H}{I} gas (same as the solid lines in Fig.~\ref{fig:all_Z_hist_med_contour}), whereas the transparent lines are the 
expected pressure dependence in thermal equilibrium, $P_{\mathrm{eq}}$, calculated as explained in Appendix~\ref{sec:thermalequilibrium} (same as the dashed lines in Fig.~\ref{fig:all_Z_hist_med_contour}) for the four ISM metallicity bins. \citet{P25} reassert that $P_{\mathrm{eq}}$ is an approximation for the thermal pressures of the simulated ISM and may be inaccurate to describe the true physical structure of the ISM, which is under the influence of dynamic processes (e.g. feedback, turbulence, shocks) not included in the static equilibrium calculations (see e.g. \citealp{Katz1996, MacLow2005}). In addition, \textsc{colibre} does not assume chemical equilibrium and has evolving dust and relative elemental abundances, which are not captured by $P_{\mathrm{eq}}$. 
The median pressures, $P_{\mathrm{median}}$, give an overview of the distribution of gas in the simulations as a function of gas density, and are therefore more connected to the simulated ISM than $P_{\mathrm{eq}}$, but they may not represent the full pressure range of cold and warm neutral gas (compare the solid lines with the contours in Fig.~\ref{fig:all_Z_hist_med_contour}). 

With these caveats in mind, Fig.~\ref{fig:rates_logn0} shows selected heating and cooling processes for gas with a density of $\log n_{\mathrm{H}}\,[\mathrm{cm}^{-3}] = 0$ \footnote{The cooling model adopts a dust-to-gas ratio of $\mathcal{DTG} = 6.6\times10^{-3}$ (dust-to-metal ratio $\mathcal{DTM} = 0.49$) and assumes chemical and thermal equilibrium and the solar relative abundances from Appendix~\ref{sec:thermalequilibrium}} for a closer analysis of the metallicity dependence in \textsc{colibre}. At this density $P_{\mathrm{median}}$ deviates from $P_{\mathrm{eq}}$ for the four metallicity bins (Fig.~\ref{fig:all_Z_Pk_T}). 
At low metallicity ($\log Z_{\mathrm{ISM}}/\mathrm{Z}_{\odot} = -1$, top left panel), the total heating (thick dashed line), and cooling (thick solid line) rates are balanced close to a temperature of $10^4\,\mathrm{K}$. Hence, the majority of the gas is expected to be in the warm phase, consistent with both $P_{\mathrm{eq}}$ and $P_{\mathrm{median}}$ (dash-dotted lines) shown in Fig.~\ref{fig:all_Z_Pk_T}. Close to the equilibrium temperature, the total cooling rate is dominated by processes related to hydrogen species, in particular collisional excitation (with subsequent Ly$\alpha$ emission) and ionization of \ion{H}{I}, as well as \ion{H}{II} recombination. Due to the reduced dust abundance at low metallicities, the dominant heating process is photo-heating of hydrogen and helium. At these low metallicities and $\log n_{\mathrm{H}}\,[\mathrm{cm}^{-3}] = 0$, the CR heating rates associated with \ion{H}{I} and \ion{He}{I} contribute significantly to the total rates. 

As metallicity increases (from left to right and top to bottom panels of Fig.~\ref{fig:rates_logn0}), PE heating on dust grains (dash-dotted line) becomes important. Note that the scaling with the dust abundance (here, with metallicity) is non-linear because an increased dust abundance also increases the self-shielding against the intrinsic radiation field, which in turn reduces PE heating. In equilibrium, gas with a metallicity of $\log Z_{\mathrm{ISM}}/\mathrm{Z}_{\odot} = -0.5$ (top-right panel), is still expected to be warm because the equilibrium temperature remains close to $10^4\,\mathrm{K}$. Comparing the total heating and cooling rates reveals that they are comparable for temperatures $\gtrsim 10^2\,\mathrm{K}$, and a small change in the dust and H$_2$ abundances results in a lower equilibrium temperature. The median pressures of the \textsc{colibre} galaxies at this metallicity (opaque dotted line in Fig.~\ref{fig:all_Z_Pk_T}) illustrate this. Here, the gas is in the cold phase at $\log n_{\mathrm{H}}\, [\mathrm{cm}^{-3}]=0$, in contrast to the expectations from $P_{\mathrm{eq}}$ (transparent dotted line). This may indicate that the dust abundance in the \textsc{colibre} live-dust model is below the assumed equilibrium value at this density, which will be discussed in detail in Sect.~\ref{sec:disc:dust}. 

For gas with solar abundances (bottom-left panel in Fig.~\ref{fig:rates_logn0}), the total cooling rate is dominated by H$_2$ cooling for temperatures of $\log T \,[\mathrm{K}]\gtrsim 2$, and by carbon cooling for lower temperatures. Here CR heating is comparable to PE heating. In our model, the equilibrium temperature is $\approx 100\,\mathrm{K}$ and the gas is therefore expected to be in the cold phase, but this critically depends on the cooling from H$_2$. In a model without molecular hydrogen cooling (e.g. as in \citealp{Wolfire1995, Wolfire2003}), the thermal equilibrium temperature may still be close to $\approx 10^4\,\mathrm{K}$, because carbon cooling (thin dashed line) does not exceed the total heating rate.
Variable dust abundances (from a live-dust model) not only affect the PE heating in a non-linear way because of the self-shielding from radiation, but also alter the carbon cooling rates through the depletion of carbon onto dust grains, leading to changes in the molecular hydrogen abundances (shielding of dissociating radiation and formation on dust grains) and therefore the H$_2$ cooling rates\footnote{While \citet{Ploeckinger2025_H2} showed that the abundance of H$_2$ may be overestimated in simulations with a reduced chemical network that does not include reactions between H$_2$ and oxygen species, the cosmic ray ionization rate used in \textsc{colibre} (from \citealp{Indriolo2015}) may be a factor of $\approx 9$ too high \citep{Obolentseva2024}. Reducing the cosmic ray rate shifts the \ion{H}{I}-H$_2$ transition to lower densities (see e.g. Fig.~10 in \citealp{P25}), which may balance the shift to higher densities when including oxygen species in the chemical network.}. We discuss the dust abundance of galaxy samples with solar metallicity in the L200m6 simulation in Sect.~\ref{sec:disc:dust} and compare to the equilibrium model. 

At high metallicity ($\log Z_{\mathrm{ISM}}/\mathrm{Z}_{\odot} = 0.5$, bottom right panel), the carbon cooling rate exceeds the total heating rate for $\log T \,[\mathrm{K}]\gtrsim 1.8$ and H$_2$ cooling plays an important role at higher temperatures. The neutral gas is expected to have low temperatures ($T_\mathrm{eq}\approx 1.8$), but $P_\mathrm{median}$ is higher than $P_{\mathrm{eq}}$ in the multiphase range (dashed lines in Fig.~\ref{fig:all_Z_Pk_T}). The large spread in stellar mass could contribute to this difference, as well as the uncharacteristically high metallicities for the stellar mass range. 

Fig.~\ref{fig:rates_logn0} demonstrates that total heating and cooling rates are comparable across large temperature ranges which makes $T_{\mathrm{eq}}$ and $P_{\mathrm{eq}}$ very sensitive to minor changes in the model, especially for gas with metallicities close to~$\mathrm{Z}_{\odot}$.

\section{Convergence with resolution}

\subsection{Number of galaxies}\label{sec:appendix:Ngalaxies}

    \begin{table} 
        \caption{The number of galaxies selected at each resolution in each metallicity bin.} 
        \centering
        \begin{tabular}{*{4}{S[table-column-width=1.4cm]}}
            \toprule
            $\log Z_\mathrm{ISM}/Z_\odot$ & ${\rm L025m5}$ & ${\rm L025m6}$ & ${\rm L025m7}$ \\
            \midrule
            0.5 & 2 & 3 & 4 \\
            0.0 & 43 & 70 & 54 \\
            -0.5 & 127 & 102 & 44 \\
            -1.0 & 75 & 49 & 15 \\
            \bottomrule
        \end{tabular} 
        \label{tab:Ngalaxies_resolution}
    \end{table}

The number of galaxies selected in each metallicity bin in the L025m7, L025m6, and L025m5 simulation are listed in Table~\ref{tab:Ngalaxies_resolution}. The median pressures discussed in Sect.~\ref{sec:resolution} are derived from these galaxy samples. The galaxy samples from the L200m6 simulation are the randomly selected 500 galaxies in each metallicity bin from Sect.~\ref{sec:metallicity_dependence}.

\subsection{Connection to the mid-plane pressure}\label{sec:appendix:resolution}
    
    In Sect.~\ref{sec:resolution}, we find that the thermal pressures in \textsc{colibre} do not converge with numerical resolution. 
    We discuss below an indication that the resolution-dependence of the thermal pressures is connected to the resolution-dependence of the mid-plane pressure. 

    The thermal and turbulent pressure must balance the mid-plane pressure, $P_{\rm mid}$, estimated by the vertical weight of the disk from stars, gas, and dark matter, assuming a multiphase ISM in dynamical equilibrium (see e.g. \citealp{Ostriker2010, KimOstrikerKim2013}). 
    A increase in the vertical scale-height of the gas disk due to larger values of the gravitational force softening at lower resolution (as suggested by \citealp{Benitez2018}) may reduce the mid-plane pressure. This could lower thermal pressures within the disk at lower resolution.
    Additionally, \citet{BenitezLlambay2026} find that the scale-height of self-gravitating components is not converged in \textsc{colibre}, suggesting the mid-plane pressure is not converged. 
    
    The mid-plane pressure of a self-gravitating disk without dark matter, $P_{\mathrm{mid}}$ from a numerical solution to hydrostatic equilibrium from \citet{Elmegreen1989} can be approximated by 
    \begin{equation}
        P_{\mathrm{mid}} \approx \frac{\pi}{2} G \Sigma_g \left [ \Sigma_g + \left (\frac{\sigma_g}{\sigma_{\star}}\right)\Sigma_{\star}\right ],
        \label{eq:mid-plane-pressures}
    \end{equation}
    
    \noindent
    with the mass surface density of gas, $\Sigma_{\mathrm{g}}$, and stars, $\Sigma_{\star}$, and their respective vertical velocity dispersions, $\sigma_{\mathrm{g}}$ and $\sigma_{\star}$. 
    
    \begin{figure}
        \centering
        \includegraphics[width=\linewidth]{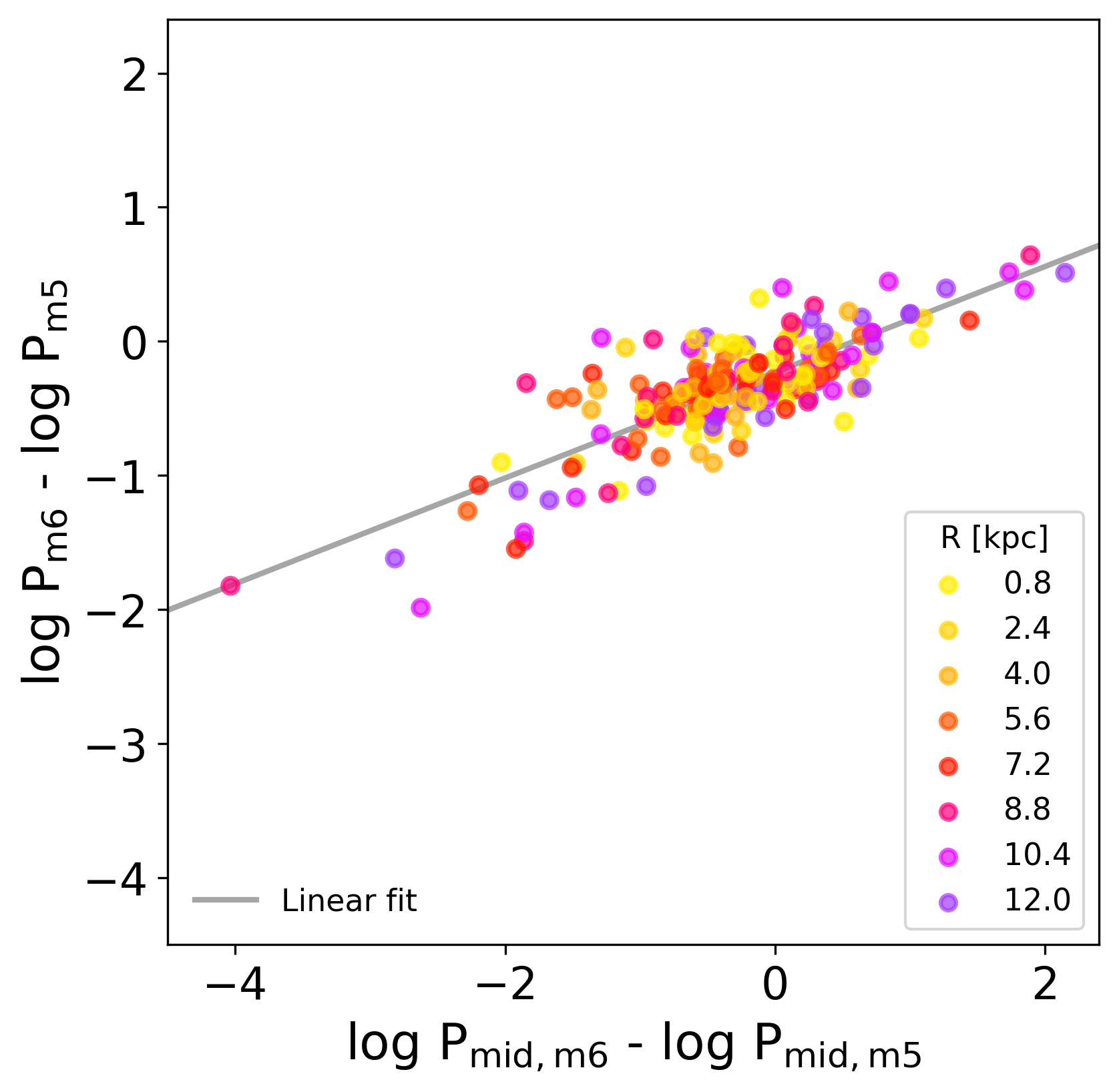}
        \caption{
            The difference in thermal pressures as a function of the difference in mid-plane pressures in radial annuli (different colours) of example galaxies with solar metallicity within the L025m6 and L025m5 simulations. The mid-plane pressures are estimated from Equation~\ref{eq:mid-plane-pressures} A simple linear fit to the data points (solid line) is shown for reference.  
        }
        \label{fig:Pmid_Pth_m6_m5_whole}
    \end{figure}
    
    We compare the mid-plane pressures (Equation~\ref{eq:mid-plane-pressures}) of all 24 galaxies with solar metallicity (from Sect.~\ref{sec:MW_analogue}) within the L025m5 simulation to their position-matched counterparts in the L025m6 simulation. Note that because of stochastic variability (e.g. \citealp{Borrow2022, Chaitra2026}) object-by-object comparisons should be avoided, since they are less robust than large-number statistics. This analysis is therefore mainly for illustration. 

    Selecting a cylindrical region extending $\pm5\, \mathrm{kpc}$ vertically above the plane of each galaxy, we calculate the thermal and mid-plane pressure in radial annuli. The expectation is a decrease in the thermal pressure from m5 to m6 resolution because the mid-plane pressure is reduced. The difference in mid-plane pressures between m6 and m5 resolution is correlated with the difference in thermal pressures (Fig.~\ref{fig:Pmid_Pth_m6_m5_whole}) with a very shallow positive slope. 
    For $\approx65\%$ of the annuli, both the mid-plane pressure and the thermal pressure are lower at m6 resolution than at the higher m5 resolution. 

    This indicates that the lower mid-plane pressure at m6 resolution may partly explain the reduced thermal pressure of the ISM at this resolution compared to m5. \cite{KimOstrikerKim2013} estimate that the thermal pressure usually is $\approx 25\%$ of the total or mid-plane pressure. Therefore, when the mid-plane pressure changes (Fig.~\ref{fig:Pmid_Pth_m6_m5_whole}) the thermal pressure adjusts by an appropriate magnitude. 
    
    We calculated the relation for the mid-plane pressure including dark matter (see Equation~B.13 in \citealp{Corbelli2025}) and reach nearly identical results. We refer to future work for a more detailed analysis of the vertical distribution of the ISM and the mid-plane pressure in \textsc{colibre} for each resolution level. 

    We conclude that the mid-plane pressure is not converged with numerical resolution. This is most likely caused by the increase in the vertical scale height due to larger gravitational force softening at lower resolution. 
    The resolution-dependence of the thermal pressure of the multiphase ISM may in part be explained by the resolution-dependence of the mid-plane pressure, because the thermal pressure is expected to adjust to changes in the mid-plane pressure.

\section{Observed thermal pressures in low-metallicity environments}\label{sec:disc:obs_lowZ} 

\begin{figure} 
    \centering
    \includegraphics[width=\linewidth]{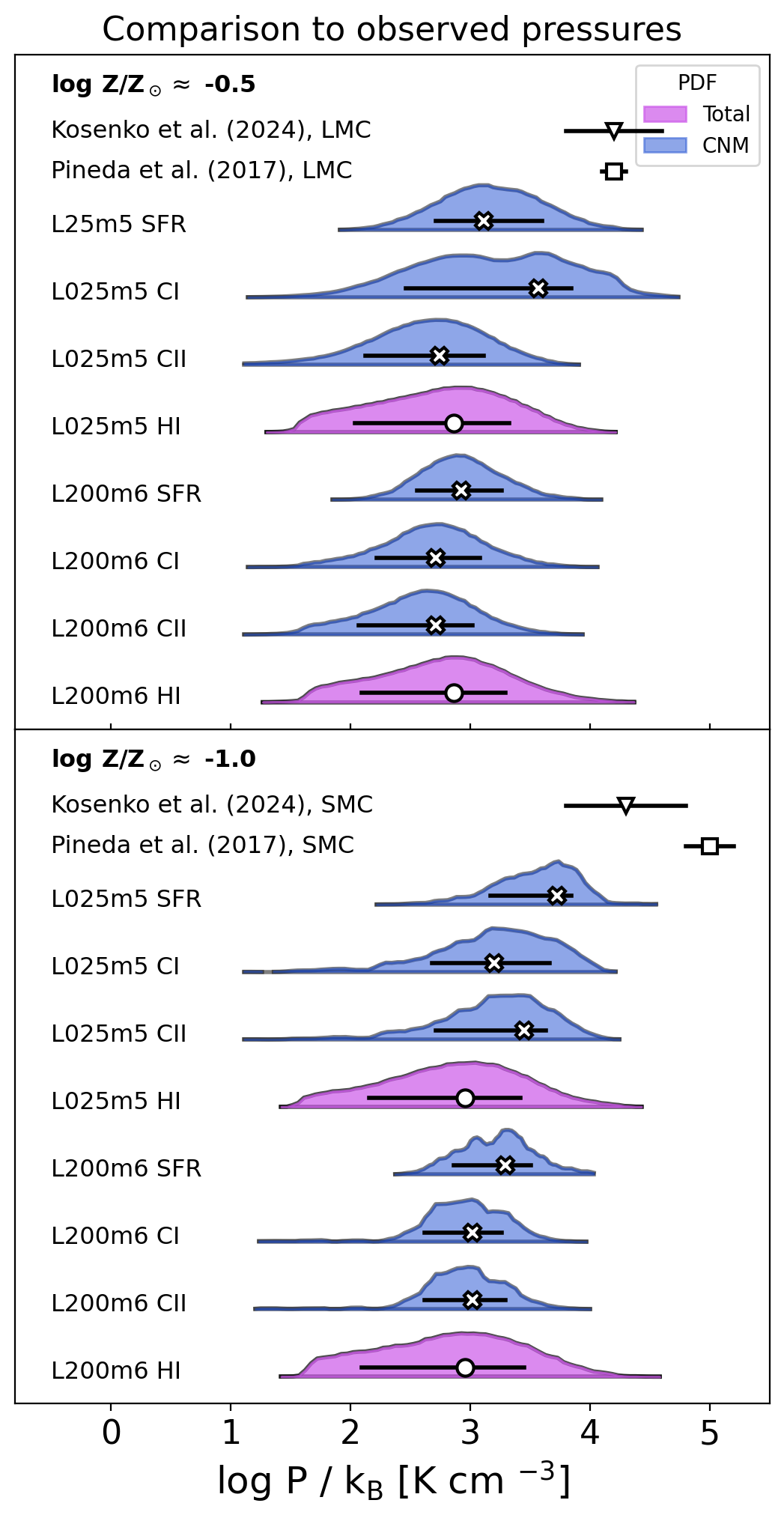}
    \caption{Thermal pressures of warm and cold neutral gas at a metallicity of $\log Z/\mathrm{Z}_{\odot} \approx -0.5$ (top panel) and -1.0 (bottom panel). 
    The observed thermal pressures in the LMC and SMC \citep{Pineda2017, Kosenko2024} compared to the thermal pressure distributions for galaxy samples, within the corresponding metallicity bin, in the L025m5 and L200m6 simulations weighted by the SFR and the \ion{C}{I}, \ion{C}{II}, and \ion{H}{I} masses of gas particles where maxima are marked by plus signs (WNM) and crosses (CNM). }
    \label{fig:thermalpressures_obs_lowerZ}
\end{figure}

In Fig.~\ref{fig:thermalpressures_obs_lowerZ}, we compare the observed thermal pressures in the LMC (middle panel) and SMC (bottom panel) to the thermal pressure distributions of galaxy samples in \textsc{colibre} simulations with mean ISM metallicity close to the metallicity of the Magellanic clouds, $\log Z_\mathrm{ISM}/Z_\odot = -0.5$ (LMC) and $-1 \pm 0.1$ (SMC). The distributions are weighted by the SFR and \ion{C}{I}, \ion{C}{II}, and \ion{H}{I} mass of gas particles in the CNM (blue) and the total neutral phase (purple) for the L200m6 and L025m5 simulations.

Representing low-metallicity environments, \citet{Pineda2017} used \textit{Herschel}/HIFI spectra of 54 sightlines toward the Small and Large Magellanic Clouds (SMC: $Z = 0.2\,\mathrm{Z}_{\odot}$, LMC: $Z=0.5\,\mathrm{Z}_{\odot}$) and report average thermal pressures of $\log P/k_{\mathrm{B}}\,[\mathrm{K\,cm}^{-3}] = 4.2$ in the LMC (excluding sightlines towards 30 Doradus) and $\log P/k_{\mathrm{B}}\,[\mathrm{K\,cm}^{-3}] = 5.0$ in the SMC, from [\ion{C}{II}] emission (Fig.~\ref{fig:thermalpressures_obs_lowerZ}). Earlier work from \citet{Welty2016}, using the \ion{C}{I} technique from \citet{JenkinsShaya1979}, included the analysis of four sightlines within the MW and 12 sightlines towards the SMC and LMC. Their MW thermal pressures are consistent with those from \citet{JT11} (see Sect.~\ref{sec:disc:obs}), while the thermal pressure measured in the Magellanic Clouds are higher, in agreement with \citet{Pineda2017}. \citet{Kosenko2024} used HST and FUSE spectra and derived thermal pressures by comparing the excitation of \ion{C}{I} fine-structure and H$_2$ rotational levels for 29 and 28 sightlines in the LMC and SMC, respectively. 
Compared to the local ISM, they found higher average thermal pressures (LMC: $\log P/k_{\mathrm{B}}\,[\mathrm{K\,cm}^{-3}] = 4.2$, SMC: $\log P/k_{\mathrm{B}}\,[\mathrm{K\,cm}^{-3}] = 4.3$) for a 5 times (2 times) stronger radiation field and 0.3~dex (0.7~dex) lower metallicity in the LMC (SMC). 

The carbon species fraction depends on the metallicity. \ion{C}{I} forms via \ion{C}{II} through recombination with free electrons or small dust grains (PAHs) \citep{Wolfire2008}, both depend on metallicity. 
Lower-metallicity environments have lower dust-to-gas ratios. The combination of a stronger FUV radiation field (like in the Magellanic Clouds) and reduced shielding by dust leads to increased photodissociation of CO. This shifts the C$^+$/C/CO transition to higher H$_2$ column densities \citep{vanDishoeckBlack1988} and increases the temperature on the outskirts of a gas cloud.  
[\ion{C}{I}] and [\ion{C}{II}] emission increases with the radiation field strength and species abundances \citep{Bisbas2021}. 
For $\log Z/Z_\odot = -1$, \citep{Bisbas2021} found that the abundance of \ion{C}{I} and \ion{C}{II} in dense clouds are equal to the CO abundance, and most carbon in \ion{H}{I}-dominated diffuse clouds is in the form of \ion{C}{II}. 
All these factors, in combination with the reasons mentioned for solar metallicity (see Sect.~\ref{sec:disc:obs}), suggest that especially \ion{C}{I} and, to a lesser extent, \ion{C}{II} trace higher densities and pressures at lower metallicities. 

The \ion{C}{I} mass at $\log Z_\mathrm{ISM}/Z_\odot = -0.5$ would populate densities of $n_\mathrm{H} \approx 2.5$ (see, e.g., Fig.~9 of \citealp{Kosenko2024}) and higher pressures, which are needed to form a stable cold phase. The L200m6 simulation is limited by resolution in this regime, while the \ion{C}{I} mass-weighted pressure distribution of the L025m5 simulation reaches higher values (Fig.~\ref{fig:thermalpressures_obs_lowerZ}). The SFR weighted pressure distribution of galaxies at this metallicity in the L025m5 simulation covers similar values to the \ion{C}{I} mass-weighted ones, but peaks at $\approx0.4~\rm dex$ lower values. 

At $\log Z_\mathrm{ISM}/Z_\odot = -1.0$, the thermal pressure distributions weighted by the SFR of galaxies in the L025m5 simulation peak at $\log P/k_{\rm B}\,[\mathrm{K\,cm}^{-3}] \approx 4$, while the \ion{C}{I} and \ion{C}{II} mass-weighted ones peak closer to $\log P/k_{\rm B}\,[\mathrm{K\,cm}^{-3}] \approx 3$ and 3.5, respectively. The fraction of \ion{C}{I} mass, and independently the fraction of \ion{C}{II} mass, in the CNM drop below $3~\%$ of the total respective mass. 

The discrepancy between observed thermal pressures in the Magellanic Clouds and the thermal pressure distributions from \textsc{colibre} galaxies (Fig.~\ref{fig:thermalpressures_obs_lowerZ}) likely arises from the strong radiation field in the LMC and SMC and the low number of sight lines, which mostly align with higher-density gas and stars. Furthermore, as seen for galaxies with solar metallicities, the dust-to-gas ratio at WNM densities from the live-dust model deviates from the equilibrium value, creating differences in the resulting heating and cooling rates from equilibrium calculation to the simulation (see Sect.~\ref{sec:disc:dust}). 

The thermal pressure distributions of the neutral phases are highly sensitive to the weighting scheme. The SFR weighted and \ion{C}{I} mass-weighted thermal pressure distributions are typically at higher pressures, comparable to the observational estimates, while the \ion{H}{I} and \ion{C}{II} mass-weighted distributions are shifted towards lower pressures.

\end{appendix}
\end{document}